\documentclass[aps,prx,twocolumn,superscriptaddress]{revtex4}
\usepackage{amsmath}
\usepackage{dcolumn}
\usepackage{amssymb}
\usepackage{amsfonts}
\usepackage{graphicx}
\usepackage[T1]{fontenc}
\usepackage{indentfirst}
\usepackage{color}
\usepackage{bbm}
\usepackage{amsbsy}
\usepackage{ulem}
\usepackage{cancel}
\usepackage[dvipsnames]{xcolor}

\begin{document}
\title{Altermagnetism and bond-nematicity in the spin-$1/2$ square lattice $J_1-J_2-\delta$ model}
\author{Tanja \DJ uri\'c}
\affiliation{School of Physical and Mathematical Sciences, Nanyang Technological University, 21 Nanyang Link, Singapore 637371}
\author{Shenlong Yu}
\affiliation{School of Physical and Mathematical Sciences, Nanyang Technological University, 21 Nanyang Link, Singapore 637371}
\author{Pinaki Sengupta}
\affiliation{School of Physical and Mathematical Sciences, Nanyang Technological University, 21 Nanyang Link, Singapore 637371}
\date{\today}
\begin{abstract}
We study appearance of bond-nematicity in insulating altermagnetic materials induced by increased frustration and quantum-fluctuations driven melting of the altermagnetic order. Using novel machine learning approach that combines symmetry enhanced neural network architectures and variational Monte Carlo we consider the spin-$1/2$ square lattice $J_1-J_2-\delta$ model known to have altermagnetic ordering in the regime of small geometric frustration and a gapless spin liquid phase in the regime of strong frustration and small exchange interaction modulation parameter $\delta$.  In the regime where exchange modulation is relatively large, resulting in the significant splitting of the magnon modes with different chiralities in the altermagnetic regime, we find that melting of the altermagnetic order by increased frustration leads to an intriguing phase that hosts coexisting symmetry protected topological valence bond solid and bond-nematic orders. The phase is characterized by condensation of magnon pairs that results in bond-nematicity, breaking of U(1) spin rotation and $\mathbb{Z}_2$ spin inversion symmetries and chiral splitting of the triplon-like energy levels in the excitation spectrum. Whilst numerous recent studies address non-trivial impact of altermagnetic moments on various properties in altermagnetic materials, like electronic band structure and superconductivity, influence of strong quantum fluctuations and phases that can result from melting of the altermagnetic order are much less explored. Our study therefore presents an important step in identifying exotic phases of matter that can emerge in vicinity of the altermagnetic order. 
\end{abstract}

\maketitle

\section{Introduction}
\label{sec:Introduction}
A newly discovered type of magnetism termed altermagnetism \cite{Smejkal, Smejkal2, Mazin, Mazin2, Jungwirth} has recently attracted considerable attention since it offers possibility of research and technology advances beyond conventional magnetism encompassing traditional ferromagnetism and antiferromagnetism. Altermagnets (AMs) uniquely combine characteristics of ferromagnets (FMs) and conventional antiferromagnets (AFMs) and exhibit a range of novel physical properties interesting both from the perspective of fundamental physics and potential device applications, particularly in the field of spintronics \cite{Jungwirth2}. 

FMs are characterized by parallel spin alignment resulting in net magnetization while in AFMs spins have antiparallel alignment canceling out net magnetization. In metallic FMs, electronic energy bands exhibit unidirectional splitting with generally momentum-independent sign corresponding to exchange-induced splitting directly related to ferromagnetic order \cite{Song}. Similarly in insulating FMs, ferromagnetic magnons are chiral \cite{Smejkal3} and can carry spin currents which is the key feature for magnon spintronics \cite{Chumak}. Contrary to electronic energy bands in metallic FMs, the energy bands in metallic AFMs do not show spin splitting in momentum space anywhere in the momentum space Brillouin zone (BZ) \cite{Song}. Similarly the opposite chirality magnon bands in insulating AFMs are degenerate across the entire BZ in the absence of an applied magnetic field \cite{Smejkal3}. This spin degeneracy of the electronic bands or chiral magnons in the whole BZ is direct consequence of the symmetries of the underlying magnetic order, in particular, those which transform sublattices with opposite spin polarization onto each other. In conventional collinear AFMs, combination of inversion or translations with time reversal leaves the magnetic order invariant leading to the spin degeneracy of electronic bands or chiral antiferromagnetic magnons. Unlike Zeeman splitting in FMs caused by non-relativistic exchange coupling  in AFMs the spin splitting of electronic bands appears only in the presence of relativistic spin-orbit coupling (SOC) such as Rashba \cite{Bychkov} or Dresselhaus \cite{Dresselhaus} SOC. 

Altermagnetism is defined by distinct class of magnetic order where, unlike for conventional AFMs, total magnetic moment does not vanish as a result of combined translations or inversion symmetries with time reversal. Instead sublattices of opposite spin polarization are transformed into each other by combination of time reversal and a point group symmetry, for example rotational symmetry. As a result AMs are characterized by higher partial wave spin structures, typically d-, g- or i-wave that generate unconventional properties of AMs like non-relativistic exchange driven spin splitting in electronic energy bands \cite{Jungwirth} in conducting AMs or splitting of magnon modes with different chiralities in insulating AMs \cite {Beida, Smejkal3, Liu,Chen2,Xie}, anomalous Hall and Nernst effects \cite{Attias,Sheoran,Yi,Weissenhofer} and piezomagnetism \cite{Ogawa, Bell}. 

Unique properties of altermagnetic materials find applications in particular in the field of spintronics \cite{Jungwirth2}. Whilst present spintronic memories are based on ferromagnetic materials with well separated and conserved spin-up and spin-down channels in the electronic structure where the magnetization in such materials limits spatial, temporal and energy scalability of such technology, altermagnetic materials offer a possibility to remove such limitations due to unique combination of well separated and conserved spin-up and spin-down channels with vanishing net magnetization.

Numerous recent studies have so far mostly addressed non-trivial impact of altermagnetic moments on various properties of altermagnetic materials, like excitation spectrum structure and superconductivity \cite{Mazin3, Liu2}. Influence of strong quantum fluctuations, introduced for example by increased geometric frustration, and possible exotic phases of matter that can emerge from melting the altermagnetic order are much less studied. An interesting recent research paper \cite{Sobral} identifies, within Schwinger-boson theory and an SU(2) gauge theory of fluctuating magnetism, possible fractionalized quantum spin liquid phases with topological order reached when quantum fluctuations destroy long-range altermagnetic spin order. In our study we identify a new unconventional disordered phase that hosts coexisting symmetry protected topological (SPT) valence bond solid (VBS) and bond-nematic (BN) orders. The phase appears as a result of frustration induced melting of the altermagnetic order within the spin-$1/2$ square lattice $J_1-J_2-\delta$ model. 

In the  $J_1-J_2-\delta$ model inequivalent surrounds of the nearest-neighboring spins, with nearest-neighbor (NN) coupling $J_1$, are mimicked by introducing $1\pm \delta$ modulation of the next-nearest-neighbor (NNN) coupling $J_2$. The model can be realized in iron oxychalcogenides \cite{Zhu} and with ultracold fermion atoms in optical lattices with a large onsite interaction \cite{Das}. Additionally spin-1 $J_1-J_2-\delta$ model could possibly be the model for monolayer V$_2$Se$_2$O and V$_2$Te$_2$O \cite{Ma}.

While altermagnetic properties of the model in the weakly frustrated regime are confirmed with the mean-field \cite{Das} and iPEPS \cite{Liu} calculations, strongly frustrated regime, in the vicinity of $J_2/J_1\approx 0.5$, is not studied. The regime features many competing phases, most prominently VBS and quantum spin liquid (QSL) phases. In the limit of vanishing modulation parameter $\delta \rightarrow 0 $ the ground state in the strongly frustrated regime is shown to be a gapless QSL \cite{Roth,Morita,Wang,Ferrari,Nomura,Liu3} confirmed in recent studies to be a $Z_2$ Dirac spin liquid \cite{Feuerpfeil,Maity}. The QSL and the neighboring VBS phase appear as Neel AFM order at small $J_2/J_1$ is replaced with the stripe AFM order at large $J_2/J_1$ \cite{Roth,Morita,Wang,Ferrari,Nomura,Liu3}. On the other hand, the $\delta \rightarrow 1$ limit corresponds to the checkerboard lattice limit with $J_2/J_1=1$ where the ground state is a plaquette valence bond solid (P-VBS) \cite{Zou}. 

We study regime of the intermediate modulation parameter $\delta$ values (around $\delta=0.5$) using recently introduced machine learning approach (ML) based on symmetry enhanced neural network architectures, in particular group equivariant convolutional neural networks (GCNNs), combined with variational Monte Carlo \cite{Roth, Duric, Duric2}. Our numerical calculations were carried out using powerful NetKet \cite{Carleo, Vicentini}, JAX \cite{Frostig}, FLAX \cite{Heek} and OPTAX \cite{Hessel} ML libraries and computational speedup on graphics processing units (GPUs). 

The GCNN based neural network quantum states (NQS) are very powerful ans\"atze that can describe complex states of matter and that in combination with stochastic reconfiguration method \cite{Sorella1,Sorella2,Rende,Chen} allow high accuracy VMC calculations. Suitability of the approach for implementation on GPUs is an additional advantage, since GPUs allow multiple parallel computations and significant computational speedup. With the GCNN and VMC approach we find a novel phase that appears  when quantum fluctuations are strong enough to melt the altermagnetic order. The phase is a SPT phase that hosts both VBS and bond-nematic orders. Our study therefore presents an important step in identifying novel phases of matter that can emerge by quantum fluctuations induced melting of the nearby altermagnetic order. Possible bond-nematic phases are particularly intriguing. In addition to fascinating interplay of symmetry and topology evident in such states of matter recent studies also show that bond-nematic phases provide possibility to simulate gravitational phenomena in laboratory settings and to study physics of linearized gravity at low energy \cite{Chojnacki}.

\section{Model and methodology}
\label{sec:Model}
\begin{figure}[t!]
\includegraphics[width=\columnwidth]{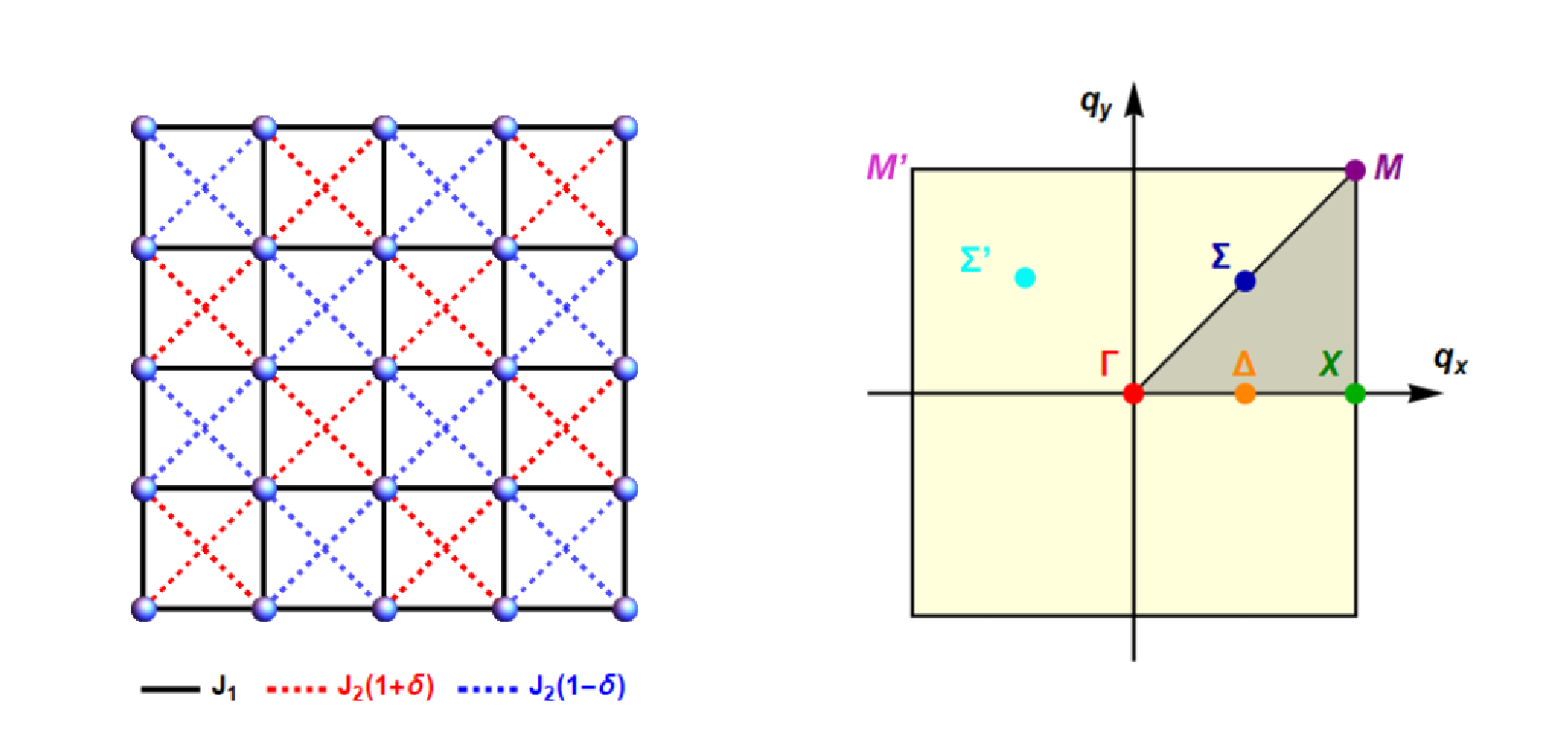}
\caption{\label{Fig:J1J2delta_BZ}Illustration of the exchange couplings in the spin-$1/2$ $J_1-J_2-\delta$ model (left panel) and the momentum space Brillouin zone (right panel) with the high symmetry points $\Gamma$, $M$ and $X$ corresponding to the crystal momenta $\vec{q}=(0,0)$, $(\pi,\pi)$ and $(\pi,0)$, respectively. Mirror reflection  over the vertical line of reflection connects $M$ and $M'$ ($\Sigma$ and $\Sigma'$) points. 
}
\end{figure}

The $J_1-J_2-\delta$ model is given by the Hamiltonian
\begin{eqnarray}\label{eq:J_1-J_2-delta_H}
H&=&\sum_{\langle i, j \rangle}J_1\hat{\vec{S}}_i\cdot\hat{\vec{S}}_j+\sum_{\langle\langle i, j \rangle\rangle}J_2(1+\delta)\hat{\vec{S}}_i\cdot\hat{\vec{S}}_j \nonumber\\ 
&+&\sum_{\langle\langle i, j \rangle\rangle'}J_2(1-\delta)\hat{\vec{S}}_i\cdot\hat{\vec{S}}_j.
\end{eqnarray}
The NN coupling is denoted by $J_1$ and NN bonds by $\langle i,j\rangle$. The NNN bonds $\langle\langle i, j \rangle\rangle$ and $\langle\langle i, j \rangle\rangle'$ have couplings $J_2(1+\delta)$ and $J_2(1-\delta)$ as illustrated in Fig. \ref{Fig:J1J2delta_BZ}. In our calculations we consider $L\times L\times 4$ clusters with straight (rectangular) edges and periodic boundary conditions (PBC) to find extrapolated results in the thermodynamic limit. We note that for such clusters unit cell contains four sites. Alternatively clusters with zig-zag edges and two atoms within the unit cell can be defined that have all symmetries of the system in the thermodynamic limit ($L\rightarrow \infty$) even for a finite size system. Since $L\times L\times 4$ clusters allow more efficient numerical calculations within the approach used in this study and described further in this section, and also allow studying larger system sizes, we consider $L\times L\times 4$ clusters to find results extrapolated to the thermodynamic limit. The thermodynamic limit results ultimately do not depend on the clusters shape.

To study properties of the ground state and magnon excitations we apply recently introduced machine learning approach based on symmetry enhanced neural network architectures combined with variational Monte Carlo (VMC) \cite{Duric,Duric2,Roth} using the neural network ansätze for the Hamiltonian eigenstates, named in literature neural network quantum states (NQS). The NQS ansatz associates a complex number $\psi(\vec{\sigma},\vec{\alpha})$ with each spin basis configuration $|\sigma_1,...,\sigma_{N_s}\rangle$:
\begin{equation}\label{eq:NQS}
|\psi\rangle= \sum_{\vec{\sigma}}\psi(\vec{\sigma};\vec{\alpha})|\vec{\sigma}\rangle
\end{equation}
where $N_s$ is the number of lattice sites and $\vec{\alpha}$ denotes the network parameters that can be found by a suitably chosen optimization algorithm. In our calculations complex coefficients $\psi(\vec{\sigma},\vec{\alpha})$ are of the GCNN form \cite{Duric,Duric2,Roth}:
\begin{equation}\label{eq:ansatz_GCNN_a}
\psi(\vec{\sigma})=\sum_{\hat{g}\in G}\chi_g^*\sum_{n=1}^{N_f}\exp\left(f^{N_l}_{n,g}\right).
\end{equation}
with $N_l$ layers, where $f^{N_l}_{n,g}$ correspond to embedding and group equivariant convolutional layers with $N_f$ complex-valued feature maps composed of two real-valued feature maps and $\hat{g}$ denotes elements of the relevant nonabelian symmetry group $G$. We set the number of layers and number of feature maps per layer to be $N_l=4$ and $N_f=4$ for the $L\times L\times 4$ clusters with $L=4,...,8$. The number of layers for the largest $9\times 9\times 4$ cluster is increased to $N_l=5$ (with $N_f=4$ features in each layer) to better capture long-range correlations. Local cluster for which elements of the embedding kernels are non-zero corresponds to the 16-sites square cluster centered at the origin for lattice sites.

A set of learnable kernels in each layer can be found by optimizing GCNN ans\"atze with suitably chosen optimization scheme. We optimized  the ans\"atze in each symmetry sector via VMC scheme combined with the natural gradient descent (NGD) method where the parameter updates in each iteration are:
\begin{equation}\label{eq:NGD_a}
\vec{\alpha}_{t+1}=\vec{\alpha}_t-\eta M^{-1} \vec{\nabla}_{\vec{\alpha}}\mathcal{L},
\end{equation}
$\mathcal{L}$ is a loss function that the VMC scheme aims to minimize and $\eta$ is the learning rate in the optimization algorithm that corresponds to the neural network training. The quantum expectation values within the optimization algorithm are estimated from $2^{12}$ samples obtained using Markov chain Monte Carlo (MCMC) sampling method within a particular fixed $S_z^{tot}$ symmetry sector and with the exchange update rule. The correlators used to calculate structure factors for the optimized NQS ans\"atze are estimated from $2^{20}$ samples. In our VMC calculations, the NGD method corresponds to the SR method \cite{Sorella1,Sorella2,Rende,Chen} where the metric matrix $M$ corresponds to the quantum
geometric tensor (QGT). \cite{Duric, Duric2, Roth} Specifically, we use the kernel (minSR) formulation of SR (NGD) \cite{Rende,Chen}, which is particularly advantageous for calculations for larger system sizes and leads to exactly the same parameter updates as the standard SR formulation. We also note that preoptimization with noisy stochastic gradient descent (SGD) \cite{Hessel} incorporated within SR, before the optimization with SR with standard SGD, significantly helps with the optimization instabilities and with avoiding high energy local minima during the neural network training. Noisy SGD is a variant of SGD that includes Gaussian noise into the updates; adding noise to the gradients can improve both the training error and the generalization error in deep neural networks.

Within the VMC scheme the loss function that is minimized is the free energy loss function \cite{Duric,Duric2,Roth}:
\begin{equation}\label{eq:free_energy}
\mathcal{L}_F\equiv F=E-TS,
\end{equation}
where a pseudoentropy reward term with an effective temperature T is added to the standard energy loss function since it encourages more even sampling of the Hilbert space and helps with
instabilities at the initial stages of the training. The effective temperature is gradually lowered to zero from an initial value $T=T_0$ as the training proceeds. In our calculations $T_n=T_0\cdot e^{-\lambda n}$ in the $n$-th training step (iteration) with $T_0=0.5$ and $\lambda=0.02$.

\begin{figure}[b!]
\includegraphics[width=\columnwidth]{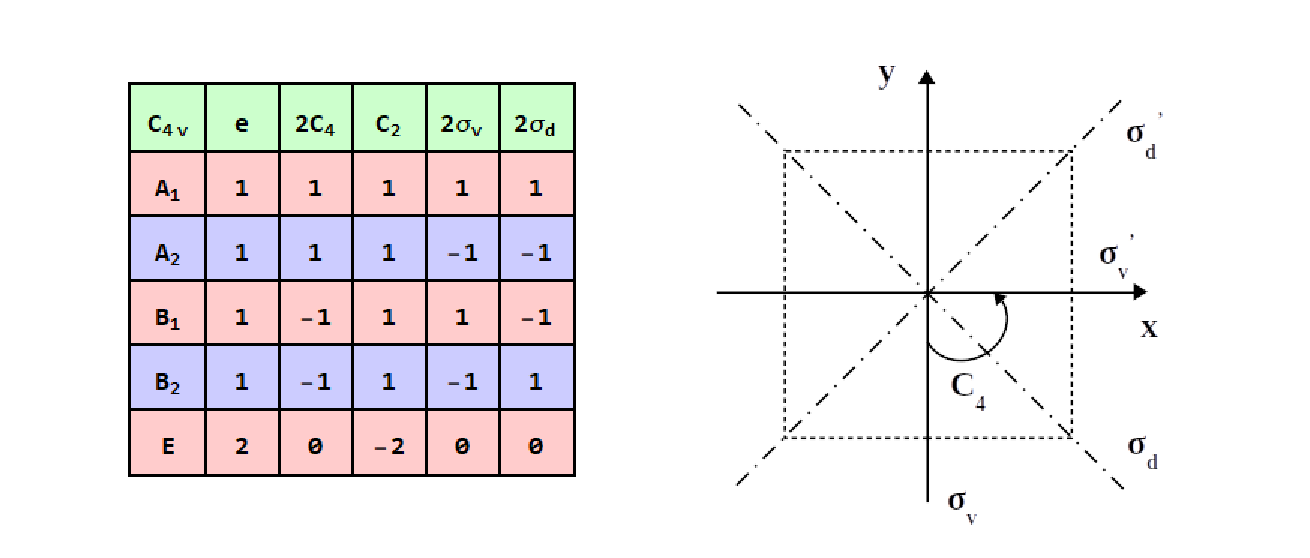}
\caption{\label{Fig:C4v}Group elements and characters of the $C_{4v}$ point group corresponding to the little group for the crystal momenta $\vec{q} =(0,0)$, $(\pi,0)$,$(\pi,\pi)$ and $(\pm \pi/2, \pi/2)$. The $C_{4v}$ group has one doubly degenerate irrep that corresponds to doubly degenerate eigenstates of the $J_1-J_2-\delta$ model defined on the $L\times L\times 4$ clusters with straight (rectangular) edges shown in Fig. \ref{Fig:J1J2delta_BZ} and with PBC. }
\end{figure}

GCNN architectures are deep neural network architectures that take into account all space-group symmetries of the Hamiltonian, namely lattice translations and point group symmetries such as rotations and reflections characterized for the $J_1-J_2-\delta$ model by $C_{4v}$ point group symmetries shown in Fig. \ref{Fig:C4v}. Since each symmetry operator $\hat{g}$ within the space group $G$ commutes with the Hamiltonian (\ref{eq:J_1-J_2-delta_H}), $\left[\hat{H},\hat{g}\right]=0$, eigenstates of the Hamiltonian correspond to irreducible representations (irreps) of the symmetry group $G$ that are eigenfunctions of all symmetry operators $\hat{g}\in G$. Each irrep has different set of eigenvalues called characters which are denoted by $\chi_g$. 

Characters that correspond to $C_{4v}$ point group are shown in Fig. \ref{Fig:C4v} for irreps $A_1$, $A_2$, $B_1$, $B_2$ and $E$. For non-degenerate irreps, characters have values $+1$ or $-1$. On the other hand, degenerate irreps, that correspond to degenerate eigenstates of the Hamiltonian, can have characters different from $\pm 1$. The full space group irreps can then be constructed by considering symmetry related crystal momenta called star and a subgroup of the point group called little group that leaves symmetry related crystal momenta invariant.
For the crystal momenta considered in our calculations, viz., $\vec{q} =(0,0)$, $(\pi,0)$, $(\pi,\pi)$ and $(\pm \pi/2, \pi/2)$, the little group that leaves crystal momenta invariant corresponds to $C_{4v}$ group and therefore we find one degenerate irrep (eigenstate), denoted by $E$ in Fig.   \ref{Fig:C4v} for all considered crystal momenta.

We note that in addition to the space-group symmetries, composed of lattice translations and $C_{4v}$ point group transformations, GCNN ans\"atze for the $M_z=S_z^{tot}=\sum_{i=1}^{N_s} S_i^z=0$ sector also take into account $\mathbb{Z}_2$ spin parity. The $\mathbb{Z}_2$ spin parity group is generated by $P_{\mathbb{Z}_2}=\prod_i\sigma_i^x$ for the eigenstates with $M_z=0$. Irreps of the spin parity group are then specified by the eigenvalues of $P_{\mathbb{Z}_2}$  that can be $P=+1$ or $P=-1$.

\section{Weak frustration regime: altermagnetism}
\label{sec:Altermagnet}
We first review properties of the system in the AM regime realized for weak frustration. We present results for $J_2/J_1=0.2$ and $\delta=0.5$ when the system is in the AM regime well away from any of the critical points where quantum phase transitions to other phases can occur. The AM regime was so far studied using the mean-field approximation  \cite{Das} and iPEPS formalism \cite{Liu}. The results showed that the altermagnetic order is reflected in chiral splitting of the magnon excitations.

\begin{figure}[t!]
\includegraphics[width=\columnwidth]{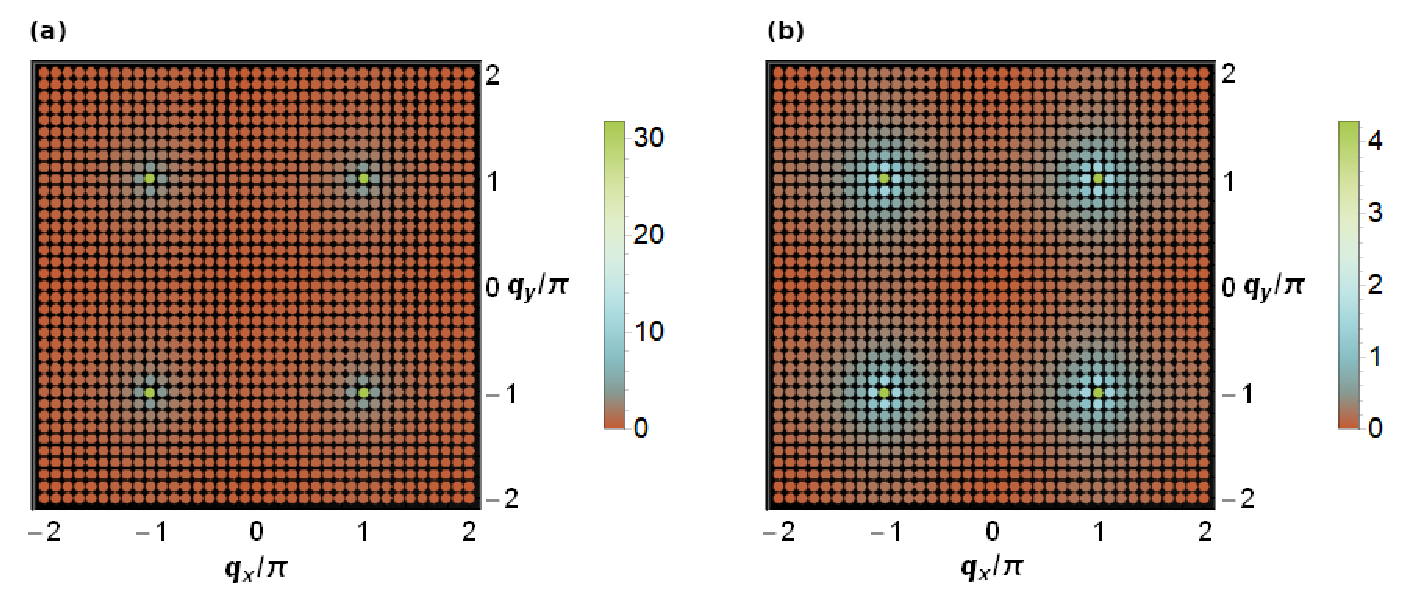}
\caption{\label{Fig:Sf_J2=0.2_9x9x4}Static spin structure factors $S_f (\vec{q})$ and $S_f^z(\vec{q})$ defined in Eqs. (\ref{eq:Sf}) and (\ref{eq:Sf_z}) for the cluster size $9\times9\times 4$ and with PBC for $J_2/J_1=0.2$ and $\delta=0.5$. The structure factors show sharp peaks corresponding to the AM order. 
}
\end{figure}

\begin{figure}[t!]
\includegraphics[width=\columnwidth]{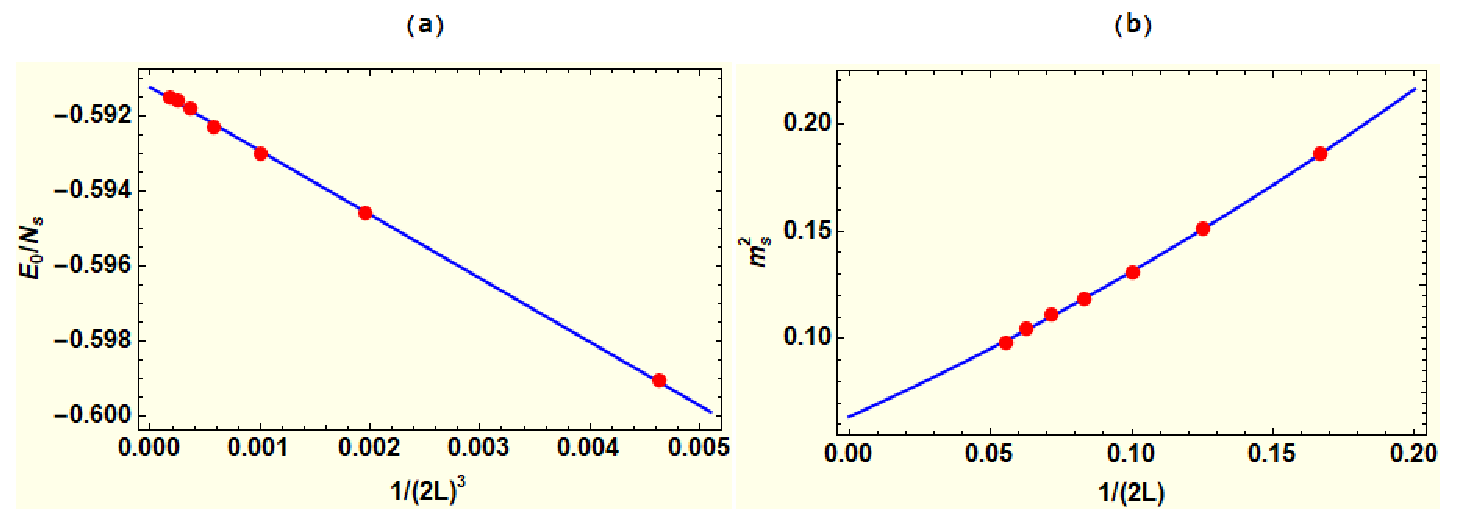}
\caption{\label{Fig:E0_M2_J2=02}The ground state energy and the squared staggered magnetization order parameter finite size scaling for $J_2/J_1=0.2$ and $\delta=0.5$. The system linear dimension is $2L$ for the $L\times L\times 4$ cluster with PBC. The ground state corresponds to the irrep $A_1$ in Fig. \ref{Fig:C4v} at the crystal momentum $\vec{q}=(0,0)$ and with the spin parity eigenvalue $P=1$ in the $S_z^{tot}=0$ sector.
}
\end{figure}

By performing optimization in each symmetry sector we find that the ground state corresponds to the irrep $A_1$ at the crystal momentum $\vec{q}=(0,0)$ and with the spin parity eigenvalue $P=1$ in the $S_z^{tot}=0$ sector. To characterize the ground state we further consider finite size scaling for the ground state energy and squared staggered magnetization order parameter. Namely, the presence of AM order can be detected by considering the thermodynamic limit value of the squared staggered magnetization defined as
\begin{equation}\label{eq:m2}
m^2_s(L)=S_f(L;\pi,\pi)/N_s
\end{equation}
where $2L$ is the lattice linear dimension, $N_s=4\times L\times L$ is the number of lattice sites for the $L\times L\times 4$ clusters that we have considered and $S_f$ is the static spin structure factor. Since AM order relates to the underlying square lattice the static spin structure factor can be defined without considering the basis for the $J_1-J_2-\delta$ model on the square lattice, that is, by taking into account only the spin-spin correlators 
\begin{equation}\label{eq:ss_cf}
C(\vec{r})=\langle \hat{\vec{S}}_0\cdot\hat{\vec{S}}_{\vec{r}}\rangle
\end{equation}
The spin structure factor then corresponds to the Fourier transform:
\begin{equation}\label{eq:Sf}
S_f(\vec{q})=\sum_{\vec{r}}e^{i\vec{k}\cdot \vec{r}} C(\vec{r}).
\end{equation}
To examine $SU(2)$ symmetry breaking we also consider the structure factor $S_f^z$ defined as a Fourier transform that corresponds only to $z$ component of the spin-spin correlators:
\begin{equation}\label{eq:Sf_z}
S_f^z(\vec{q})=\sum_{\vec{r}}e^{i\vec{k}\cdot \vec{r}} C^z(\vec{r}).
\end{equation}
where $C^z(\vec{r})=\langle \hat{S}_0^z\cdot\hat{S}_{\vec{r}}^z\rangle$. The structure factors exhibit sharp peaks corresponding to the AM order. As an example, the results for the $9\times 9\times 4$ cluster are shown in Fig. \ref{Fig:Sf_J2=0.2_9x9x4}. The ground state energy per site and squared staggered magnetization order parameter scaling is shown in Fig. \ref{Fig:E0_M2_J2=02}. 

Extrapolated thermodynamic value of the ground state energy per lattice site $E_0(L\rightarrow \infty)\approx -0.5912$ is in good agreement with the iPEPS result \cite{Liu}. The extrapolated thermodynamic value of the squared staggered magnetization $m_s^2(L\rightarrow \infty)\approx 0.06364$ ($\sqrt(m_s^2)\approx 0.2523$) however is somewhat smaller than the iPEPS prediction \cite{Liu}, possibly due to limited bond dimension or optimization algorithm in the iPEPS calculations. NQS ML formalism can often capture quantum fluctuations more efficiently which results in stronger renormalization of the magnetization order parameter that is more strongly suppressed by quantum fluctuations \cite{Viteritti}.

In the AM regime the ground state energy per site $e_0(L)=E_0(L)/N_s$ scales asymptotically with the lattice linear dimension $2L$ as:
\begin{equation}\label{eq:E0_scaling}
e_0(L)\approx e_0(\infty) + A/{L}^3,
\end{equation}
equivalent to the scaling of the Heisenberg AFM on the square lattice. In the presence of the magnetic order (AM or AFM) the squared magnetization in general scales as:
\begin{equation}\label{eq:M2_scaling}
m_s^2(L)\approx m_s^2(\infty) +A_1/L+A_2/{L}^2. 
\end{equation}
We also note that NQS ans\"atze break $SU(2)$ spin rotation symmetry even for the finite size systems \cite{Moss}. The $SU(2)$ symmetry breaking can be clearly seen for larger system sizes by calculating the structure factor $S_f^z$. For an $SU(2)$ symmetric state $S_f^z=S_f/3$. The results shown in Fig. \ref{Fig:Sf_J2=0.2_9x9x4} clearly demonstrate that $S_f^z\neq S_f/3$ indicating that the NQS and VMC algorithm finds a state that includes the ground state and portion of the low-lying excited states that collapses onto the ground state in the thermodynamic limit causing spontaneous $SU(2)$ symmetry breaking and appearance of the altermagnetic order, and not the true finite size system ground state that is $SU(2)$ symmetric. 

In our opinion, in the NQS and VMC calculations, the variational NQS ansatz attempts to capture broken symmetry state and mimic thermodynamic limit behavior \cite{Moss}, causing large energy variance for the lowest energy NQS state for finite size clusters if the spontaneous symmetry breaking happens in the thermodynamic limit. For phases with magnetic order, relevant low-lying excited states are those within the Anderson tower of states \cite{Anderson, Tasaki, Wietek}. However, since the Anderson tower of states is responsible for the spontaneous symmetry breaking in the thermodynamic limit, scaling to the thermodynamic limit results in the correct value of the order parameter \cite{Moss}. 

To further confirm that the ground state has altermagnetic and not antiferromagnetic ordering we study lowest energy states in each symmetry sector that correspond to chiral magnon modes. Unlike AFMs, AMs exhibit splitting between magnon modes with opposite chiralities \cite{Beida, Smejkal3, Liu,Chen2,Xie}. For the $J_1-J_2-\delta$ model the linear spin-wave theory (LSWT) gives two branches of magnon modes \cite{Liu,Toth}:
\begin{equation}\label{eq:LSWT}
\omega^{\pm}(\vec{q})=\frac{\sqrt{c_2^2-4c_1^2ˇ}}{2} \pm \frac{c_3}{2},
\end{equation}
where $c_1=J_1(\cos(q_x)+\cos(q_y))$, $c_2=4J_1-4J_2(1-\cos(q_x)\cos(q_y))$ and $c_3=4J_2\delta\sin(q_x)\sin(q_y)$. For nonzero $c_3$ (nonzero modulation parameter $\delta$) the magnon spectrum is split to two branches $\omega^+(\vec{q})$ and $\omega^-(\vec{q})$ with the maximum splitting at $\vec{q}=(\pm \pi/2,\pm \pi/2)$ reflecting $d_{xy}$ symmetry of the splitting. For vanishing modulation parameter $\delta$ there is no splitting between two chiral modes that are therefore degenerate indicating antiferromagnetic order, as is expected in that limit.

Although LSWT is strictly justified only in the large-$S$ limit with minimal quantum fluctuations, it captures qualitatively altermagnetic splitting of the chiral magnon modes. The LSWT however cannot provide the results that are quantitatively comparable with the experimental observations since it neglects magnon-magnon and other interactions important in strongly correlated systems. Further calculations were therefore done within the iPEPS formalism \cite{Liu} to obtain more accurate magnon spectrum for the $J_1-J_2-\delta$ model. Apart from the Nambu-Goldstone mode the iPEPS calculations find much more accurate magnon spectrum. Difficulty to capture Nambu-Goldstone mode, gapless magnon excitation that appears because of the spontaneous $SU(2)$ symmetry breaking, stems from the finite bond dimension effect and can be improved by increasing bond dimension within the iPEPS algorithm.

\begin{figure}[t!]
\includegraphics[width=\columnwidth]{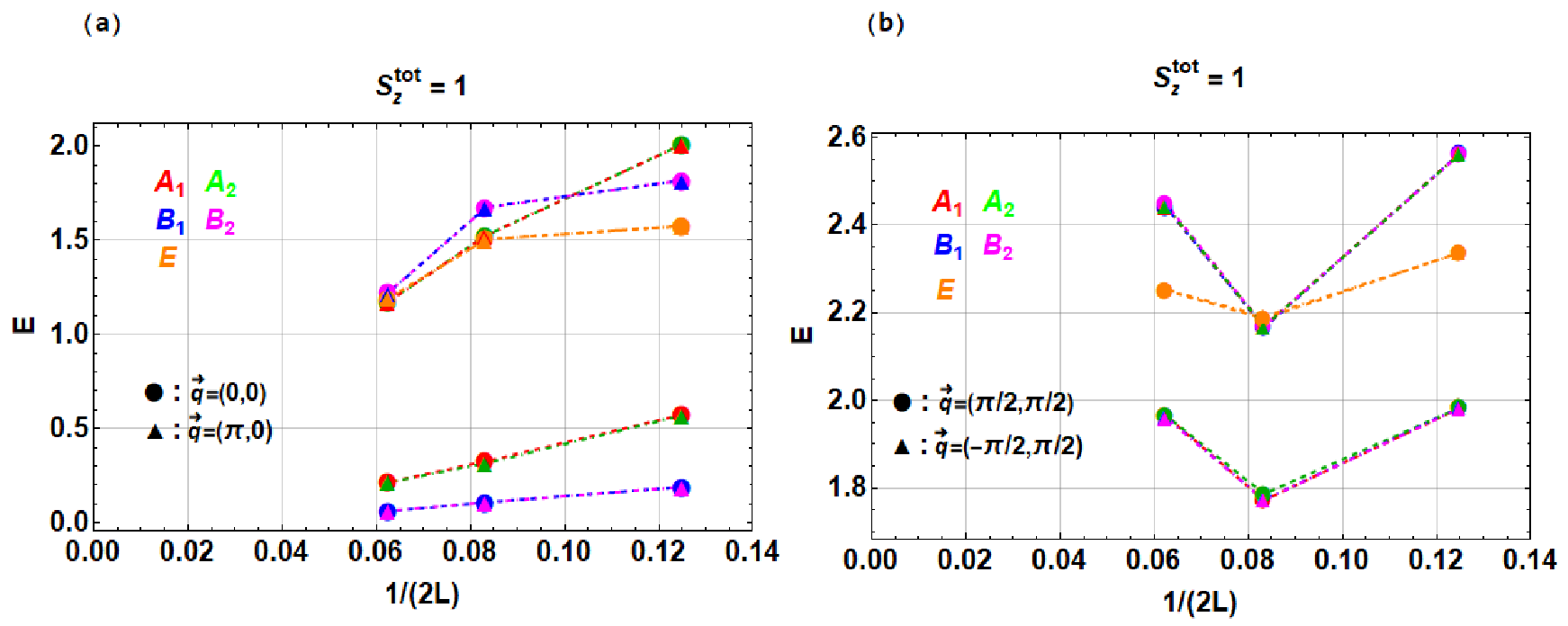}
\caption{\label{Fig:E_splitting}Chiral magnon splitting in the altermagnetic regime with $J_2/J_1=0.2$ and $\delta=0.5$ for the crystal momenta (a) $\vec{q}=(0,0)$, $(\pi,0)$ ($\Gamma$ and $X$ high symmetry points in Fig. \ref{Fig:J1J2delta_BZ}) and (b) $\vec{q}=(\pm\pi/2,\pi/2)$ ($\Sigma$ and $\Sigma'$ high symmetry points in Fig. \ref{Fig:J1J2delta_BZ}). $A_1$, $A_2$, $B_1$, $B_2$ and $E$ denote irreps of the $C_{4v}$ point group that correspond to the eigenstates in different symmetry sectors. Energy levels for $\vec{q}=(\pi,\pi)$ ($M$ high symmetry point) are identical to the energy levels for $\vec{q}=(0,0)$. As explained in the main text the magnon modes corresponding to the irreps $A_1$ and $B_1$, and equivalently to the irreps $A_2$ and $B_2$ have opposite chiralities. Also, $A_1$ and $A_2$ / $B_1$ and $B_2$  magnon modes have opposite chiralities. 
}
\end{figure}

Our GCNN and VMC results for the lowest energy magnon modes within each symmetry sector described by irreps $A_1$, $A_2$, $B_1$, $B_2$ and $E$ in Fig. \ref{Fig:C4v} and at crystal momenta $\vec{q}=(0,0)$, $(\pi,0)$, and $(\pm \pi/2, \pi/2)$ are shown in Fig. \ref{Fig:E_splitting} (energy spectrum at $\vec{q}=(\pi,\pi)$ is identical to the energy spectrum at $\vec{q}=(0,0)$). The two lowest energy magnon modes with opposite chiralities correspond to the irreps $A_1$ and $B_1$ at the crystal momentum $\vec{q}=(0,0)$ (and $\vec{q}=(\pi,\pi)$) and to the irreps $A_2$ and $B_2$ at the crystal momentum $\vec{q}=(\pi,0)$ as demonstrated in Fig. \ref{Fig:E_splitting} (a). Namely, for an altermagnetic state with $d_{x^2-y^2}$ symmetry, as it is for the $J_1-J_2-\delta$ model, $C_4$ rotation and $\sigma_d$ diagonal mirror reflection symmetry operations map one sublattice to another while simultaneously flipping the spin direction. Irreps $A_1$ and $B_1$ with different characters transform differently with respect to $C_4$ rotation and diagonal mirror reflection $\sigma_d$,  as shown in the character table in Fig. \ref{Fig:C4v}. They correspond to (quasi)degenerate magnon modes with opposite chiralities at high-symmetry points. The same argument is valid for irreps $A_2$ and $B_2$. 

The maximum splitting between the modes corresponding to irreps $A_1$ and $B_1$, (equivalently between $A_2$ and $B_2$) can be seen by comparing energy levels at $\vec{q}=(\pi/2,\pi/2)$ and $\vec{q}=(-\pi/2,\pi/2)$ as shown in Fig. \ref{Fig:E_splitting} (b), since changing $\vec{q}=(\pi/2,\pi/2)\rightarrow (-\pi/2,\pi/2)$ changes chirality of the modes. The degenerate irrep $E$ corresponds to linear combination of two modes with opposite chiralities and therefore does not show chiral splitting along high-symmetry directions. We note that $A_1$ magnon mode represents in-phase oscillation of sublattice spins, while $B_1$ magnon mode represents out-of-phase oscillations. Accordingly, $A_1$ is a symmetric s-wave type mode with clockwise (right-handed) chirality, $B_1$ mode is an antisymmetric d-wave type mode with counter-clockwise (left-handed) chirality that transforms under $C_{4v}$ symmetry operations as $d_{x^2-y^2}$ orbital.

We further note that irreps $A_1$ and $A_2$ also have opposite chiralities and are energetically split, as do irreps $B_1$ and $B_2$. The modes corresponding to irreps $A_1$ and $A_2$ ($B_1$ and $B_2$) however show maximum splitting at $\vec{q}=(0,0)$ and $(\pi,0)$ and are degenerate at $\vec{q}=(\pm\pi/2,\pi/2)$. This is evident from Fig. \ref{Fig:E_splitting} by comparing energy levels at the crystal momenta $\vec{q}=(0,0)$ and $\vec{q}=(\pi,0)$ in panel (a) and energy levels at $\vec{q}=(\pm\pi/2,\pi/2)$ in panel (b). Changing crystal momentum from $\vec{q}=(0,0)$ to $\vec{q}=(\pi,0)$ ($(0,\pi)$) corresponds to the translation by one lattice site along $x$ ($y$) direction. The lattice translation by one lattice site causes exchange of the local magnetic environments that causes exchange of chiral characteristics between the modes. The two lowest energy magnon modes with the crystal momentum $\vec{q}=(\pi,0)$ therefore correspond to the irreps $B_2$ and $A_2$ and not to the irreps $B_1$ and $A_1$.

Fig. \ref{Fig:E_splitting} (a) also clearly demonstrates that the chiral splitting between two lowest energy magnon modes with opposite chirality decreases with the increase of the system size and vanishes in the thermodynamic limit at the high symmetry points $\Gamma$, $M$ and $X$ shown in Fig. \ref{Fig:J1J2delta_BZ}. Energy gap for the magnon excitations at the high symmetry points $\Gamma$, $M$ and $X$ also decreases with the increase of the system size and vanishes in the thermodynamic limit. While this result agrees with the iPEPS result at $\vec{q}=(0,0)$, at the crystal momentum $\vec{q}=(\pi,0)$ iPEPS calculations identified only the roton minimum that corresponds to the degenerate $A_1$ and $B_1$ irreps in the thermodynamic limit at $\vec{q}=(\pi,0)$ \cite{Liu}. If the quantum fluctuations are taken into account more accurately and efficiently roton minimum energy decreases. This can already be seen by comparing LSWT and iPEPS results \cite{Liu}. Our GCNN and VMC approach finds lower energy for the roton minimum than iPEPS calculations \cite{Liu} and also identifies two magnon modes with energies below the roton minimum that correspond to the (quasi)degenerate $A_2$ and $B_2$ irreps. Energy gap for the magnon excitations at $\vec{q}=(\pi,0)$  therefore vanishes in the thermodynamic limit. We also note that the energy levels for $S_z^{tot}=1$ and $S_z^{tot}=-1$, and $\vec{q}$ and $-\vec{q}$, are identical. 

Fig. \ref{Fig:E_splitting} (b) clearly shows chiral splitting at the crystal momentum $\vec{q}=(\pm\pi/2,\pi/2)$ ($\Sigma$ and $\Sigma'$ points in Fig. \ref{Fig:J1J2delta_BZ}) where LSWT and iPEPS calculations find the maximum splitting. We argue that the eigenstates that correspond to irreps $E$, $B_1$ and $B_2$ are degenerate in the thermodynamic limit at the crystal momentum $\vec{q}=(\pi/2,\pi/2)$, and equivalently irreps $E$, $A_1$ and $A_2$ at the crystal momenta $\vec{q}=(\pi/2,\pi/2)$. The energy levels for the $L\times L\times 4$ clusters at $\vec{q}=(\pm\pi/2,\pi/2)$ would then correspond to the energy levels of the clusters with zig-zag edges and two sites in the unit cell at $\vec{q}=(\pm\pi/2,\pi/2)$ in the thermodynamic limit. For the clusters with zig-zag edges the eigenstates at $\vec{q}=(\pm \pi/2,\pi/2)$ are described by two irreps of the $C_s$ group that consists of identity and reflection with respect to the diagonal reflection axis that closes $\pi/4$ angle with the $x$-axis. Two irreps have characters $[1,1]$ or $[1,-1]$. Since the energy spectrum for both types of clusters with PBC should be the same in the thermodynamic limit the energy levels $E$, $B_1$ and $B_2$ ($E$, $A_1$ and $A_2$) will exhibit degeneracy and evolve into one energy level as the system size is increased. The degenerate energy level $E$ does not show chiral splitting because it is composed of two modes with opposite chiralities. Taking into consideration degeneracies in the thermodynamic limit our results show good agreement with the iPEPS results at the point of maximum chiral splitting \cite{Liu}.

\section{Strong frustration regime: Bond nematicity and SPT VBS order} 
%and SPT valence bond solid and bond-nematic orders in the highly frustrated regime.}
 \label{sec:VBS_nematic}

We further study magnetically disordered regime that appears when frustration is increased and quantum fluctuations destroy altermagnetic order. We consider regime of relatively large modulation parameter $\delta$ for $J_2/J_1=0.5$ that is within the range where frustration is maximal. In our calculations we set modulation parameter $\delta=0.5$. In the limit $\delta \rightarrow 0$ the ground state is a gapless spin liquid \cite{Roth,Morita,Wang,Ferrari,Nomura,Liu3} found in recent studies to be a $Z_2$ Dirac spin liquid \cite{Feuerpfeil,Maity}. For relatively large values of delta, however, we find the ground state exhibits simultaneous SPT VBS and bond-nematic orders. The quantum phase transition between such phase and $Z_2$ Dirac spin liquid could possibly correspond to a deconfined quantum critical (DQC) point \cite{Senthil, Senthil2, Cui}.  

\begin{figure}[t!]
\includegraphics[width=\columnwidth]{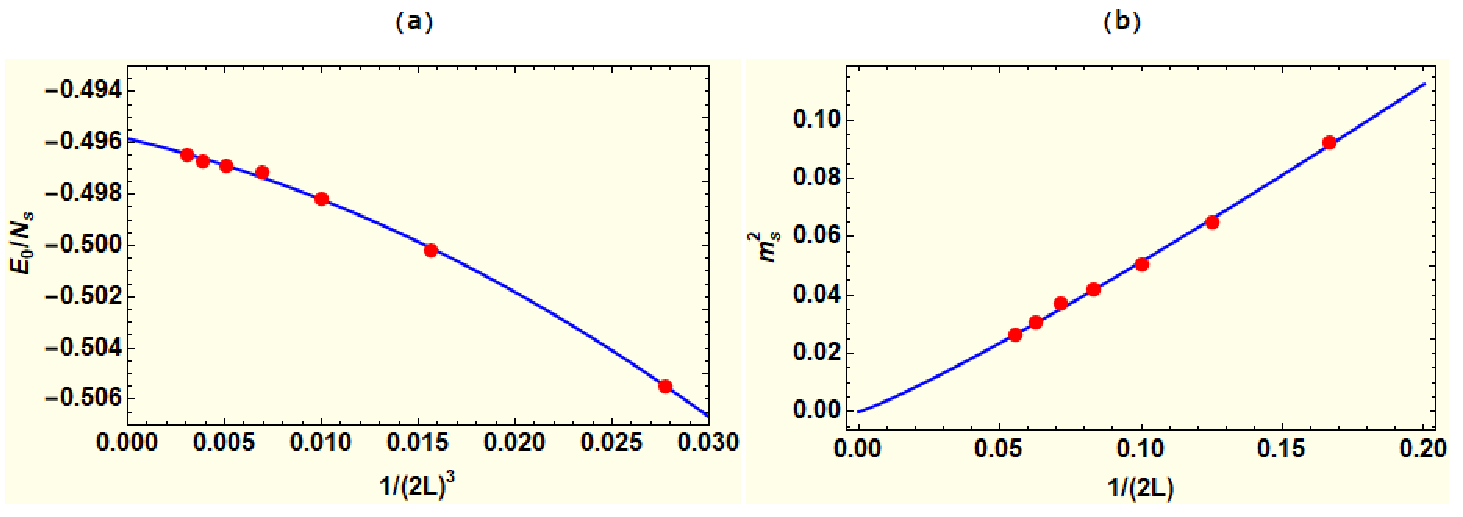}
\caption{\label{Fig:E0_M2_J2=05}The ground state energy and the squared staggered magnetization order parameter finite size scaling for $J_2/J_1=0.5$ and $\delta=0.5$. The system linear dimension is $2L$ for the $L\times L\times 4$ cluster with PBC. Unlike in the AM regime the squared staggered magnetization order parameter shows power-law behavior and vanishes in the thermodynamic limit. The ground state corresponds to the irreps $A_1$ in Fig. \ref{Fig:C4v} at the crystal momentum $\vec{q}=(0,0)$ and with the spin parity eigenvalue $P=1$ in the $S_z^{tot}=0$ sector.
}
\end{figure} 
To characterize the phase in the highly frustrated regime, we first examine the ground state energy and  staggered magnetization. The results for finite-sized systems are shown in Fig. \ref{Fig:E0_M2_J2=05}, along with their scaling to the thermodynamic limit. 

\begin{figure}[b!]
\includegraphics[width=\columnwidth]{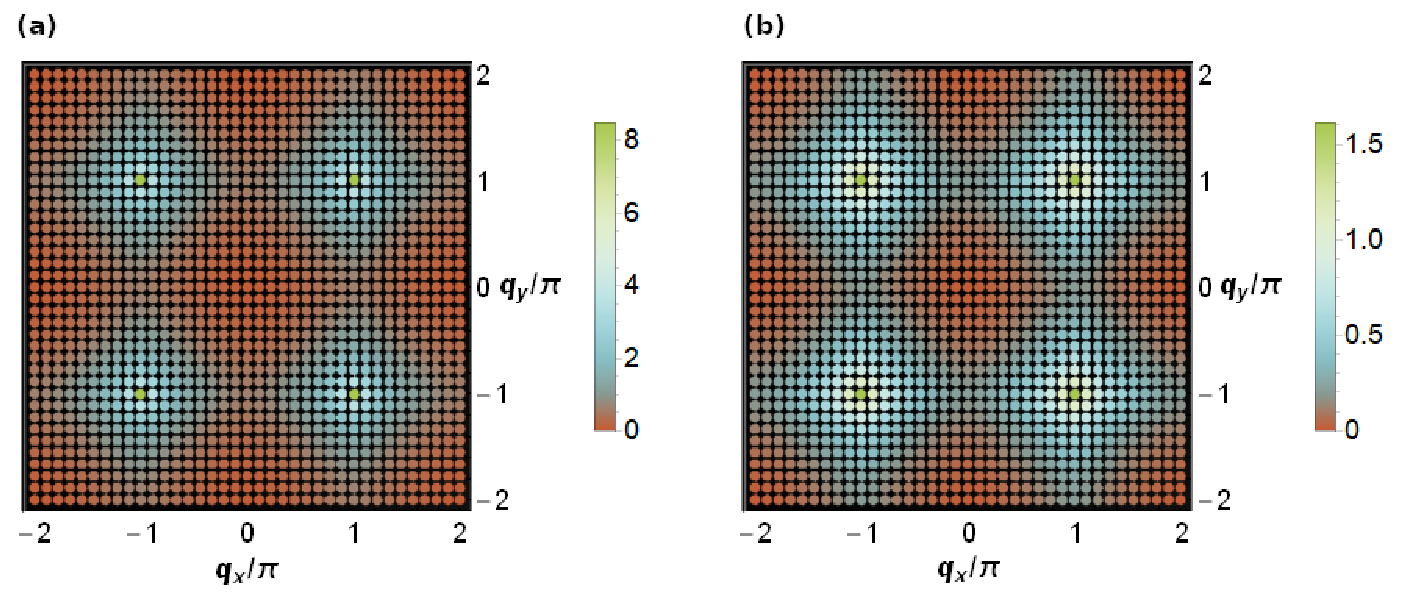}
\caption{\label{Fig:Sf_J2=0.5_9x9x4}Static spin structure factors $S_f (\vec{q})$ and $S_f^z(\vec{q})$ defined in Eqs. (\ref{eq:Sf}) and (\ref{eq:Sf_z}) for the cluster size $9\times9\times 4$ and with PBC for $J_2/J_1=0.5$ and $\delta=0.5$. The structure factors show much broader peaks than in AM regime (Fig. \ref{Fig:Sf_J2=0.2_9x9x4}) signaling disordered phase without AM order. 
}
\end{figure}

The ground state energy per site for $J_2/J_1=0.5$ scales as:
\begin{equation}\label{eq:E0_scaling_J2=0.5}
e_0(L)\approx e_0(\infty) + B_1/{L}^3 +B_2/{L}^6,
\end{equation}
where $e_0=E_0/N_s$. Different scaling of the ground state energy than in the AM regime is evident and signals a different phase. The squared staggered magnetization order parameter shows critical scaling:
\begin{equation}\label{eq:M2_scaling_J2=0.5}
m_s^2(L)\approx B\cdot {L}^{-(1+\eta)},
\end{equation}
with $\eta \approx 0.126$ and vanishes in the thermodynamic limit, $m_s^2(\infty)=0$, demonstrating that the ground state in the highly frustrated regime is magnetically disordered. The vanishing of magnetic ordering is corroborated by  spin structure factors (defined in Eqs. (\ref{eq:Sf}) and  (\ref{eq:Sf_z})) that show much broader peaks at $\vec{q}=(\pi,\pi)$ indicating disappearance of the altermagnetic order (Fig. \ref{Fig:Sf_J2=0.5_9x9x4}).
%as shown for the $9\times 9\times 4$ cluster in  

We further find that the lowest energy NQS state still significantly breaks $SU(2)$ spin rotation symmetry, and more prominently as the system size is increased, although there is no $SU(2)$ symmetry breaking in finite systems. 
%due to the altermagnetic ordering. 
The energy has large variance that increases with the system size. This indicates that $U(1)$ spin rotation symmetry is also broken in the thermodynamic limit, in addition to the $SU(2)$ symmetry. As we will demonstrate further in this section the $U(1)$ symmetry is broken due to magnon pair condensation that leads to a bond-nematic order \cite{Zhitomirsky, Wang3}. For the bond-nematic ordered phases the low energy excitations form bond-nematic equivalent of the Anderson tower of states \cite{Yokoyama,Momoi} and collapse of such states to the ground state in the thermodynamic limit causes spontaneous symmetry breaking. In addition, the ground state has SPT VBS order \cite{Haghshenas,Wang2} characterized by a finite dimer order parameter in the thermodynamic limit and different behavior of the scaling for the linear dimensions of the system $2L=4n$ and $2L=4n+2$ corresponding to trivial and topological (SPT) sectors. Our calculations for the lowest energy excitations, presented further in this section, additionally reveal the ground state degeneracy consistent with the columnar-like order that breaks $C_4$ rotational symmetry, rather than plaquette-like VBS order that maintains $C_4$ and rotation symmetries. In particular, we find that the nematic Goldstone modes, that are a direct consequence of the ground state spontaneous symmetry breaking, break $C_4$ rotation and reflections symmetries implying that the same symmetries are broken in the ground state.

Bond-nematicity is normally defined in terms of the traceless rank two tensor \cite{Shannon} :
\begin{equation}\label{eq:N_op}
\hat{O}^{\alpha\beta}(\vec{r}_i,\vec{r}_j)=(\hat{S}_i^\alpha\hat{S}_j^\beta +\hat{S}_i^\beta\hat{S}_j^\alpha)/2 -\delta^{\alpha\beta}\langle \hat{\vec{S}}_i\cdot \hat{\vec{S}}_j\rangle/3
\end{equation}
with $\alpha, \beta \in \{x,y,z\}$.
%that exhibits long range order. 
The tensor reveals hidden order by projecting onto a symmetrized spin-1 Hilbert space of bond variables with long range correlations; it is related to the magnon pairing operator via relation $\hat{S}_i^-\hat{S}_j^-=\hat{O}^{xx}-\hat{O}^{yy}-2i\hat{O}^{xy}$ and $\hat{S}_i^+\hat{S}_j^+=(\hat{S}_i^-\hat{S}_j^-)^\dagger$. Condensation of such bound pairs of magnons leads to a bond-nematic order, breaking of the U(1) spin rotation symmetry and appearance of the nematic Anderson tower of states. For spin-$1/2$ systems bond-nematicity usually appears in the presence of competing ferromagnetic and antiferromagnetic couplings \cite{Shannon} where bond-nematic correlations often have $d$-wave symmetry and corresponding structure factors are defined taking into account such $d$-wave symmetry.

\begin{figure}[b!]
\includegraphics[width=\columnwidth]{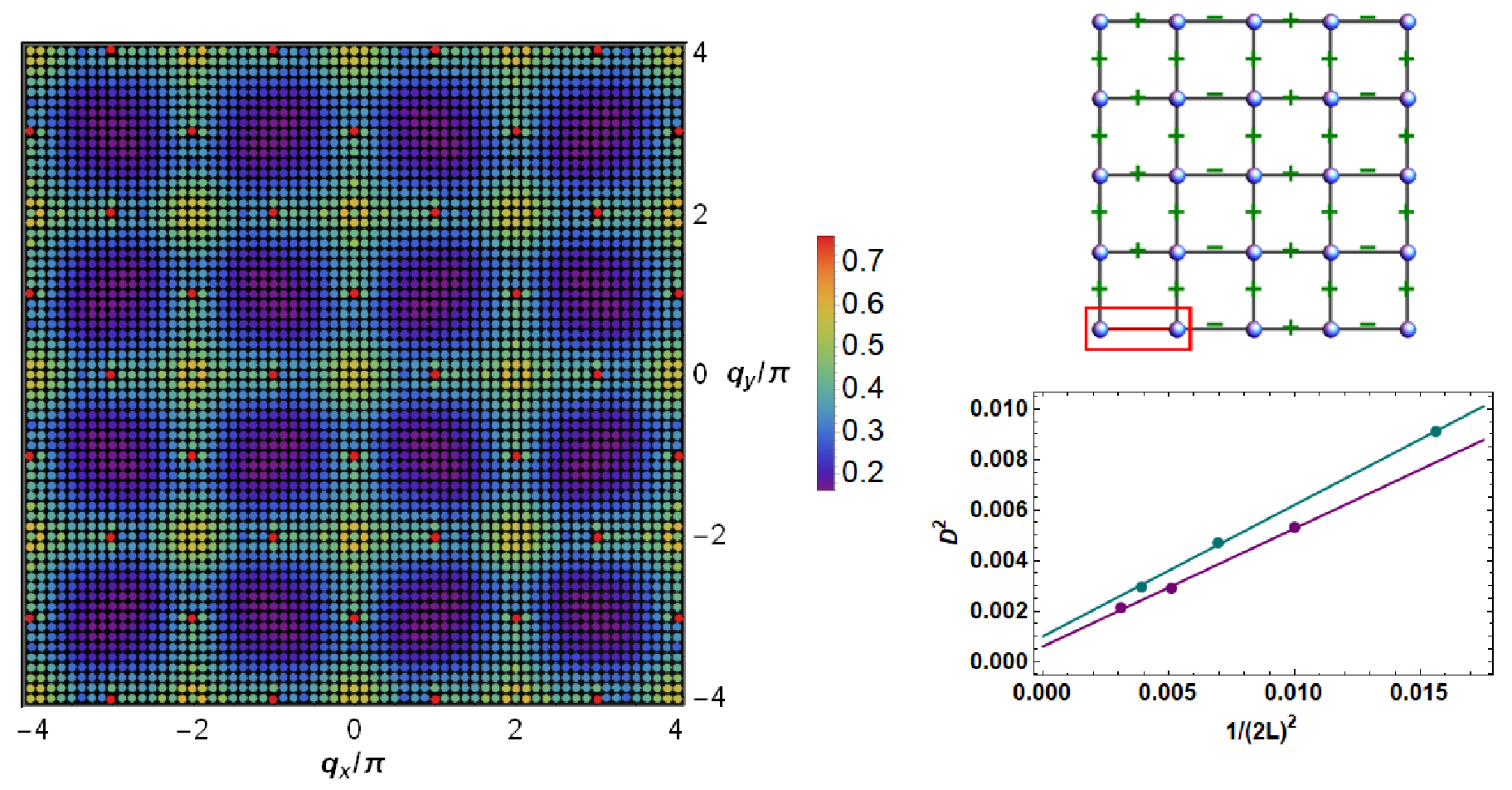}
\caption{\label{Fig:D_sf}Left panel: Dimer structure factor defined in Eq. (\ref{eq:D_sf_1}) for the $8\times 8\times 4$ cluster with PBC. Right panel: Sign structure of dimer-dimer correlators (where the red rectangle indicates the reference bond) and finite size scaling for dimer order parameter (defined in Eq. (\ref{eq:D_op_1})) obtained from $L\times L\times 4$ clusters with $2L=4n$ (cyan filled circles) and $2L=4n+2$ (magenta filled circles). 
}
\end{figure}

To define suitable correlators for the magnetically disordered phase that we have found in our calculations we first consider sign pattern for prototypical bond-nematic correlation functions $\sum_{\alpha\beta} \langle O^{\alpha\beta}(0,\vec{e}_i) O^{\alpha\beta}(\vec{r},\vec{r}+\vec{e}_j)\rangle$ for the lowest energy state, where $\vec{e}_i,\vec{e}_j\in\{\hat{x},\hat{y}\}$ and $\hat{x}$ and $\hat{y}$ are unit vectors connecting two neighboring sites in the $x$- and $y$- directions. The sign pattern is shown in the right panel of Fig. \ref{Fig:D_sf}. As it can be seen from the figure the bond-nematic correlations do not exhibit $d$-wave symmetry characterized by positive sign of correlators on parallel bonds and negative correlations on perpendicular bonds as found in spin-$1/2$ systems with competing ferromagnetic and antiferromagnetic couplings \cite{Shannon}. Instead, the bond-nematic correlations show $s$-wave symmetry with superimposed  sign modulation due to the coexisting VBS order. We therefore define the bond-nematic structure factor in terms of the following correlators:
\begin{equation}\label{eq:N_cf2}
C_N^{\alpha\beta}(\vec{r})=\sum_{\vec{e}_i,\vec{e}_j\in\{\hat{x},\hat{y}\}}C_N^{\alpha\beta}(\vec{r},\vec{e}_i,\vec{e}_j)
\end{equation}
where $\vec{e}_i,\vec{e}_j\in\{\hat{x},\hat{y}\}$ and 
\begin{equation}\label{eq:N_cf}
C_N^{\alpha\beta}(\vec{r},\vec{e}_i,\vec{e}_j)=\hat O^{\alpha\beta} (0,\vec{e}_i)\hat O^{\alpha\beta}(\vec{r},\vec{r}+\vec{e}_j),
\end{equation}
and not in terms of the correlators $\bar{C}_N^{\alpha\beta}(\vec{r})=\sum_{\vec{e}_i,\vec{e}_j\in\{\hat{x},\hat{y}\}}(-1)^{(1-\delta_{\vec{e}_i,\vec{e}_j})}C_N^{\alpha\beta}(\vec{r},\vec{e}_i,\vec{e}_j)$ that would be suitable for $d$-wave symmetry. The corresponding nematic structure factors are then defined as :
\begin{equation}\label{eq:N_f}
N_f^{\alpha\beta}(\vec{q})=\sum_{\vec{r}}e^{i\vec{k}\cdot\vec{r}}C_N^{\alpha\beta}(\vec{r})
\end{equation}
and reflect appearance of the bond-nematic order as a peak in the structure factors at the $\vec{q}=(0,0)$ crystal momentum. Additional peaks at $\vec{q}=(0,\pi)$ and $\vec{q}=(\pi,0)$ signal coexisting VBS order.

\begin{figure}[t!]
\includegraphics[width=\columnwidth]{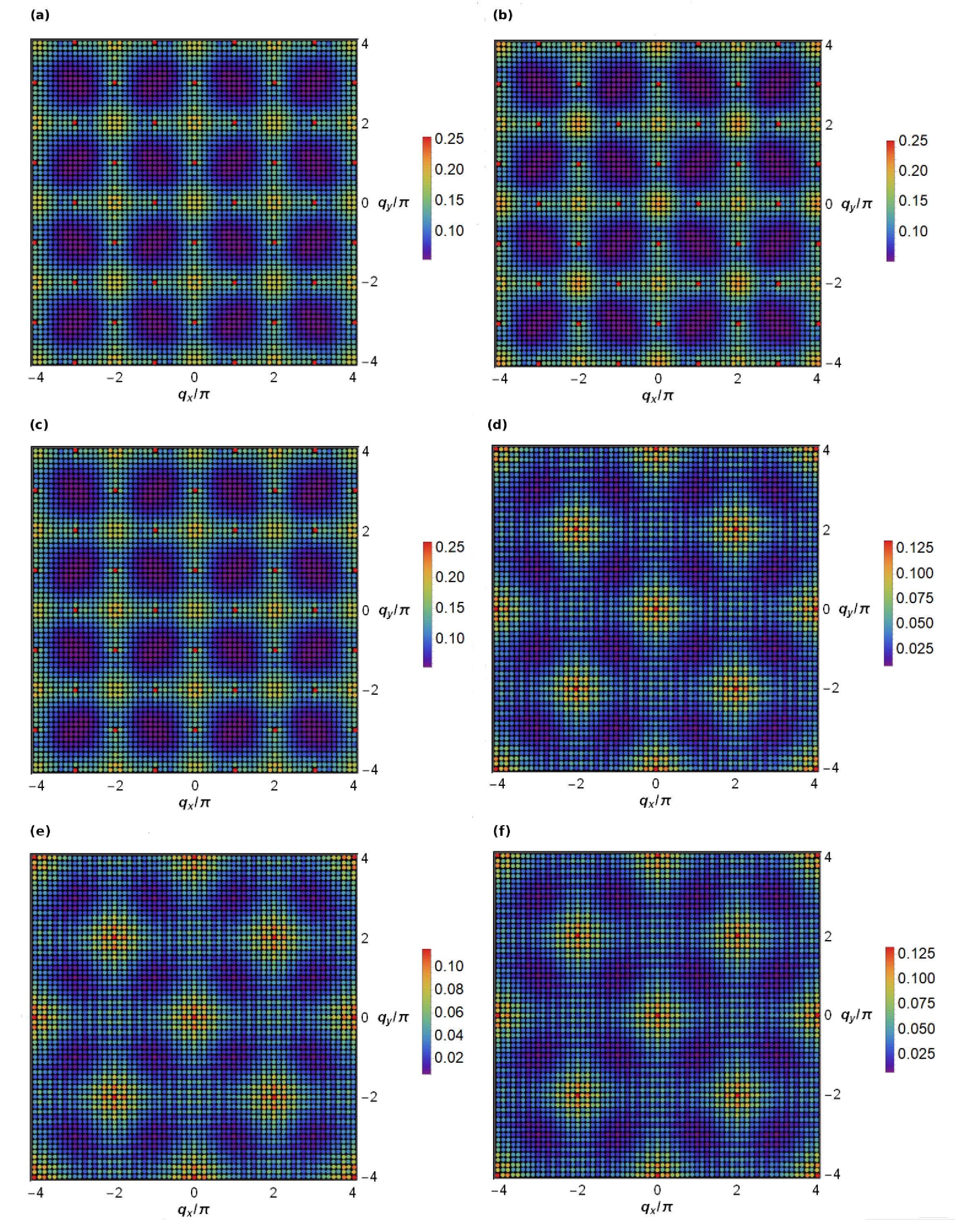}
\caption{\label{Fig:N_sf}Nematic structure factors (a) $N_f^{xx}(\vec{q})$, (b) $N_f^{yy}(\vec{q})$, (c) $N_f^{zz}(\vec{q})$, (d) $N_f^{xy}(\vec{q})$, (e) $N_f^{xz}(\vec{q})$ and (f) $N_f^{yz}(\vec{q})$ defined in Eq. (\ref{eq:N_f}) for the $8\times 8\times 4$ cluster with PBC, $J_2/J_1=0.5$ and $\delta=0.5$. Broad peak at $\vec{q} = (0,0)$ signals nematic order and sharp peaks at $\vec{q} = (\pi,0)$ and $\vec{q} = (0,\pi)$ coexisting VBS order. 
}
\end{figure}

\begin{figure}[t!]
\includegraphics[width=\columnwidth]{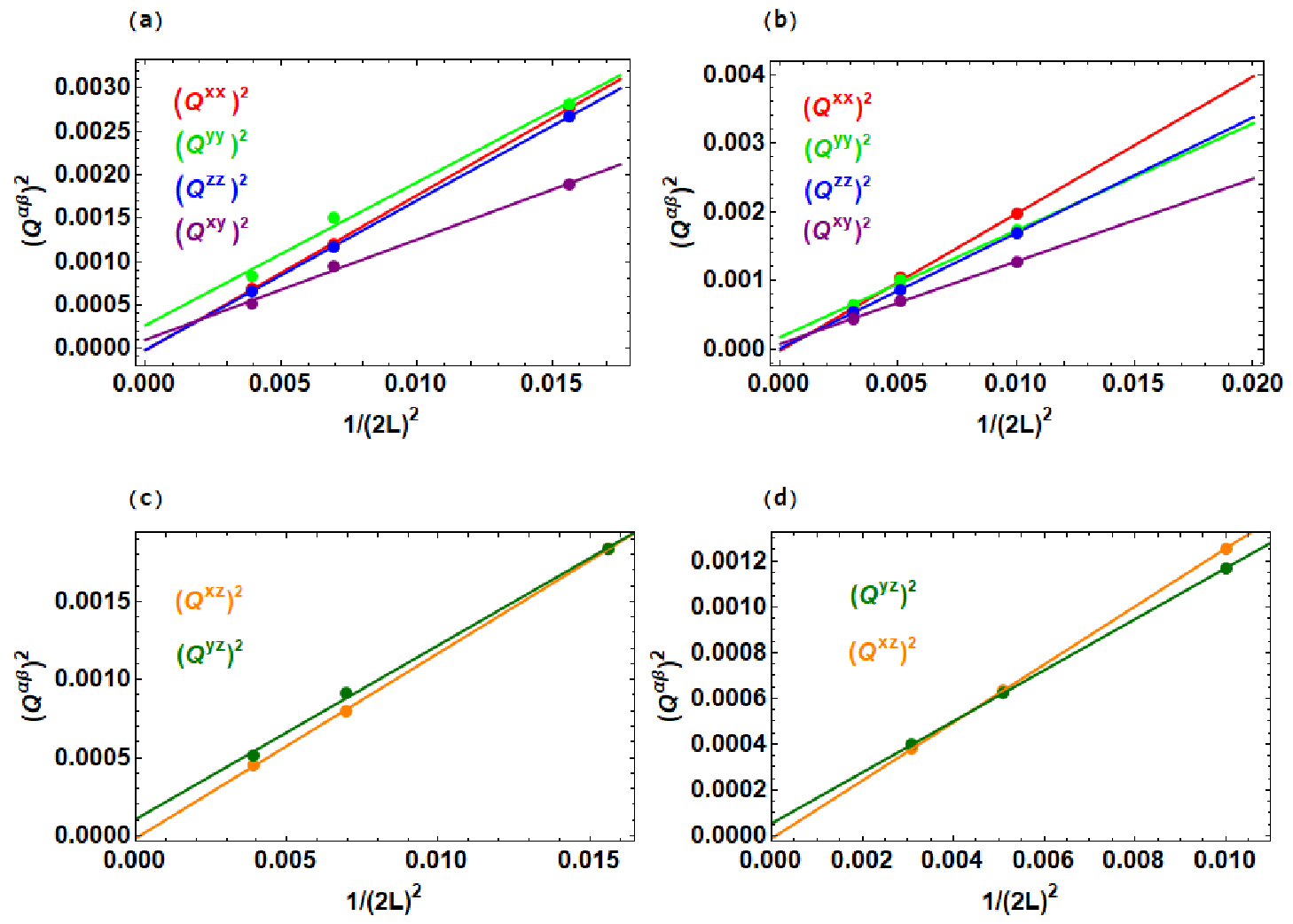}
\caption{\label{Fig:Q_scaling}Finite size scaling for squared bond-nematic order parameters $(O^{xx})^2$, $(O^{yy})^2$, $(O^{zz})^2$, $(O^{xy})^2$ (panels (a), (b)) and $(O^{xz})^2$, $(O^{yz})^2$  (panels (c), (d)) for $2L=4n$ (panels (a), (c)) and $2L=4n+2$ (panels (b), (d)). The magnon pairing operator is related to the bond-nematic order parameters via equations $\hat{S}_i^-\hat{S}_j^-=\hat{O}^{xx}-\hat{O}^{yy}-2i\hat{O}^{xy}$ and $\hat{S}_i^+\hat{S}_j^+=(\hat{S}_i^-\hat{S}_j^-)^\dagger$.
}
\end{figure}

Results for nematic structure factors for the $8\times 8\times 4$ cluster with PBC are shown in Fig. \ref{Fig:N_sf} and results for the nematic order parameters scaling in Fig. \ref{Fig:Q_scaling}. We find that the squared bond-nematic order parameters $(O^{\alpha,\beta})^2=N_f^{\alpha\beta}(\vec{q}=(0,0))/N_s$ scale as:
\begin{equation}\label{eq:O_scaling}
(O^{\alpha\beta})^2(L)\approx (O^{\alpha\beta})^2(\infty)+A_{\alpha\beta}/{L}^2,
\end{equation}
for the $L\times L\times 4$ clusters. 
The results show different scaling of the order parameters for clusters with the linear dimensions $2L=4n$ and $2L=4n+2$. As it will be explained further in this section this indicates existence of a SPT order with two distinct sectors. While $(O^{xx})^2$ and $(O^{zz})^2$ scale to zero in the thermodynamic limit, $(O^{yy})^2$ and $(O^{xy})^2$ scale to a finite value. Extrapolation to the thermodynamic limit for clusters with $2L=4n$ gives $(O^{yy})^2\approx 0.000264$ ($Q^{yy}=\sqrt{(O^{yy})^2}\approx 0.0162$) and $(O^{xy})^2\approx 0.0000967$ ($Q^{xy}=\sqrt{(O^{xy})^2}\approx 0.00983$), while for the clusters with $2L=4n+2$  $(O^{yy})^2\approx 0.000172 $ ($Q^{yy}=\sqrt{(O^{yy})^2}\approx 0.0131$) and $(O^{xy})^2\approx 0.0000735$ ($Q^{xy}=\sqrt{(O^{xy})^2}\approx 0.00857$). This confirms existence of a magnon pair condensate since $\hat{S}_i^-\hat{S}_j^-=\hat{O}^{xx}-\hat{O}^{yy}-2i\hat{O}^{xy}$. We note that condensed magnon pairs have $S_z^{tot}=0$ and $S^{tot}=2$. Energy levels with $S_z^{tot}=\pm 2$ are much higher in the energy spectrum.

Additionally the bond-nematic order parameter $(Q^{yz})^2$ also scales to a nonzero value in the thermodynamic limit, while $(Q^{xz})^2$ vanishes when $L\rightarrow \infty$. For $2L=4n$ when $n\rightarrow \infty$ $(Q^{yz})^2\approx 0.000104$ ($\sqrt{(Q^{yz})^2}\approx 0.01$) and for $2L=4n+2$ when $n\rightarrow \infty$ $(Q^{yz})^2\approx 0.0000532$ ($\sqrt{(Q^{yz})^2}\approx 0.00729$). As it will be explained further in this section nonzero $(Q^{yz})^2$ signals broken $\mathbb{Z}_2$ spin inversion symmetry.

Sharp peaks in the bond-nematic structure factors $N_f^{\alpha\alpha}(\vec{q})$ at $\vec{q}=(\pi,0)$ and $(0,\pi)$ motivates us to  define dimer structure factor $\bar{D}_f(\vec{q})$ and squared dimer order parameter $\bar{D}^2$ as:
\begin{equation}\label{eq:D_sf_1}
\bar{D}_f(\vec{q})=\sum_{\alpha\in\{x,y,z\}}  N_f^{\alpha\alpha}(\vec{q})
\end{equation}
and 
\begin{equation}\label{eq:D_op_1}
\bar{D}^2=[\bar{D}_f(\vec{q}=(\pi,0))+\bar{D}_f(\vec{q}=(0,\pi))]/(2N_s).
\end{equation}
The results for the dimer structure factor $\bar{D}_f(\vec{q})$ for $8\times 8\times 4$ cluster are shown in the left panel of Fig. \ref{Fig:D_sf}; corresponding order parameter scaling is shown in the right panel. 
%of Fig. \ref{Fig:D_sf}. 
The squared dimer order parameter $\bar{D}^2$ also scales as:
\begin{equation}\label{eq:D_scaling_1}
\bar{D}^2(L)\approx \bar{D}^2(\infty)+A_{\bar{D}}/{L}^2.
\end{equation}
Similar to  the bond-nematic order parameters, the results for the dimer order parameter show different scaling for the clusters with the linear dimensions $2L=4n$ and $2L=4n+2$. This is a characteristic feature of a SPT VBS order with $2L=4n$ and $2L=4n+2$ corresponding to a trivial and topological (SPT) sectors.

For $2L=4n$ bond inversion leads to a topologically equivalent VBS pattern because two patterns are transmutable to each other via translation. For $2L=4n+2$ bond inversion leads to a different VBS pattern that cannot be shifted to its complement without breaking a bond. The ground state is then a linear superposition of two topologically distinct states with different bond patterns which suppresses VBS order and results in a smaller value for the VBS order parameter. The VBS order parameter suppression is clearly visible in Fig. \ref{Fig:D_sf}. Extrapolation to the thermodynamic limit results in different values of the dimer order parameter for the clusters with $2L=4n$ and $2L=4n+2$, $\bar{D}^2_{4n}(n\rightarrow \infty)\approx 0.001$ and $\bar{D}^2_{4n+2}(n\rightarrow \infty)\approx 0.0006$.

\begin{figure}[b!]
\includegraphics[width=\columnwidth]{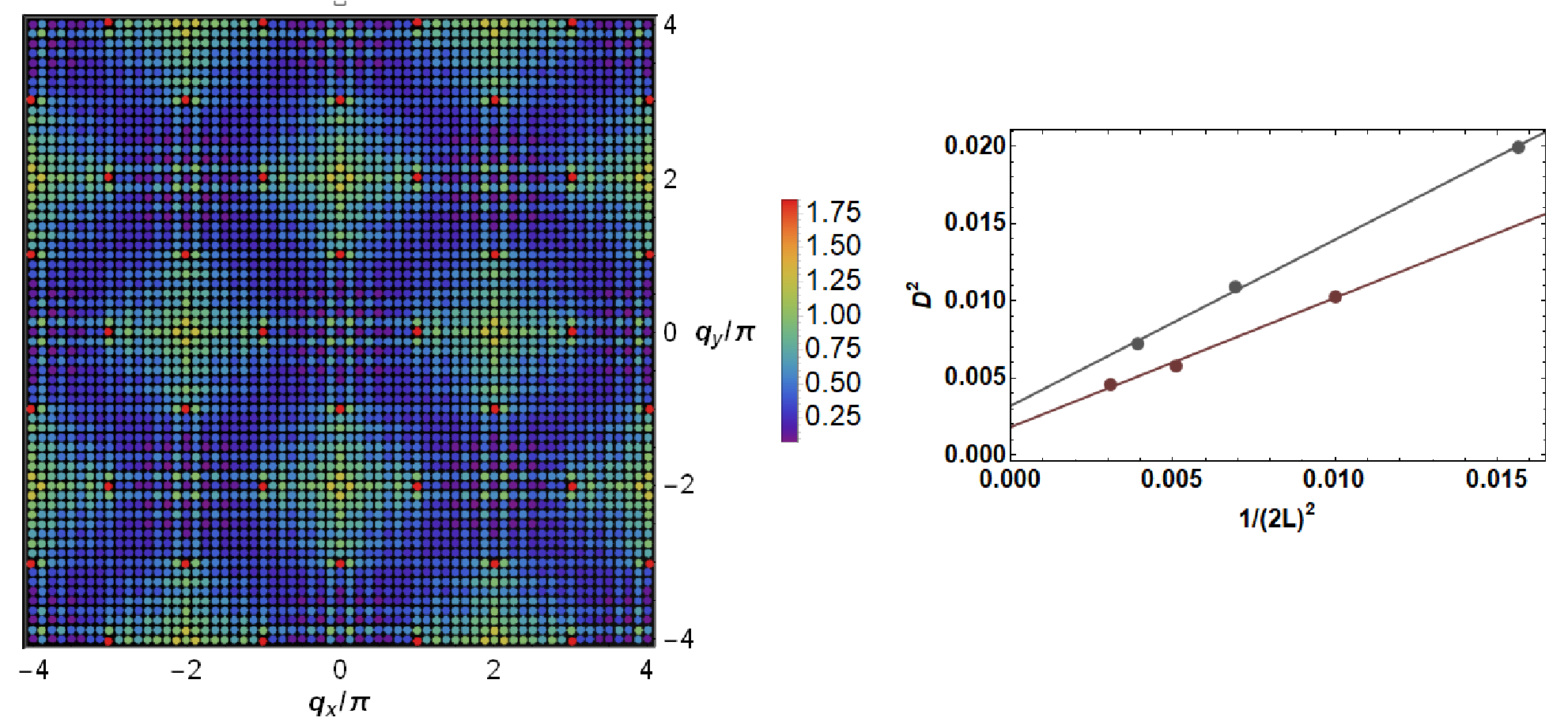}
\caption{\label{Fig:D2_II} Left panel: Dimer structure factor defined in Eq. (\ref{eq:Df}) for the $8\times 8\times 4$ cluster with PBC. Right panel: Finite size scaling for dimer order parameter (defined in Eq. (\ref{eq:Df})) obtained from $L\times L\times 4$ clusters with $2L=4n$ (gray filled circles) and $2L=4n+2$ (dark pink filled circles). 
}
\end{figure}
To verify the presence of the SPT VBS order independently from the bond-nematic order, we have also calculated dimer order parameter using standard definitions for dimer-dimer correlators 
\begin{equation}\label{eq:dd_correlators}
 C^D(\vec{r}) =\sum_{\vec{e}_i,\vec{e}_j\in\{\hat{x},\hat{y}\}}  [\langle \hat{\vec{S}}_0\cdot \hat{\vec{S}}_{\vec{e}_i} \hat{\vec{S}}_{\vec{r}}\cdot \hat{\vec{S}}_{\vec{r}+\vec{e}_j} \rangle -\langle \hat{\vec{S}}_0\cdot \hat{\vec{S}}_{\vec{e}_i} \rangle \langle\hat{\vec{S}}_{\vec{r}}\cdot \hat{\vec{S}}_{\vec{r}+\vec{e}_j} \rangle]
\end{equation}
and corresponding dimer structure factor
\begin{equation}\label{eq:Df}
 D_f({\vec{q}})=\sum_{\vec{r}} e^{i\vec{k}\cdot\vec{r}}C^D(\vec{r}). 
\end{equation}
where dimer order parameter can again be calculated as $D^2=[D_f(\pi,0)+D_f(0,\pi)]/(2N_s)$. The dimer structure factor for the $8\times 8\times 4$ cluster with PBC and the finite size scalings for the $L\times L\times 4$ clusters with PBC and $2L=4n$ and $2L=4n+2$ are shown in Fig. \ref{Fig:D2_II}. Equivalently to the order parameter $\bar{D}^2$, dimer order parameter $D^2$ also scales as: 
\begin{equation}\label{eq:D_scaling_2}
D^2(L)\approx D^2(\infty)+A_{D}/{L}^2,
\end{equation}
and exhibits different scaling behavior for the clusters with $2L=4n$ and $2L=4n+2$. Extrapolated thermodynamic limit values are $D^2_{4n}(n\rightarrow \infty) \approx 0.0032$ and $D^2_{4n+2}(n\rightarrow \infty) \approx 0.0018$.

We finally note that the spin chiral order parameter $\chi=\hat{\vec{S}}_i\cdot(\hat{\vec{S}}_j\times\hat{\vec{S}}_k)$ is zero for all triangles within the square plaquettes containing sites $i$, $j$ and $k$.

\begin{figure}[t!]
\includegraphics[width=\columnwidth]{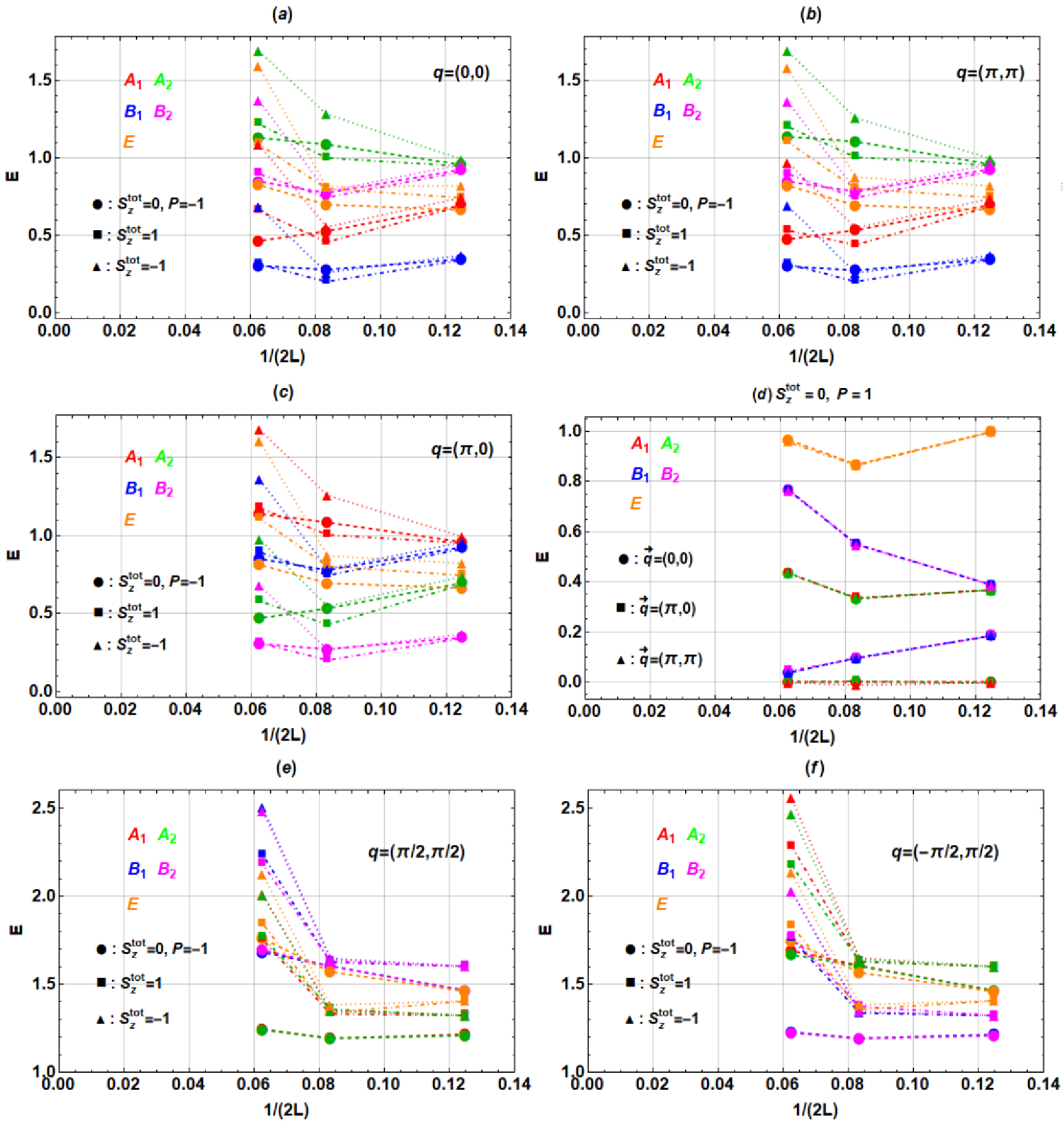}
\caption{\label{Fig:En} Lowest energy excitations in each $C_{4v}$ point group symmetry sector (irreps $A_1$, $A_2$, $B_1$, $B_2$ and $E$ in Fig. \ref{Fig:C4v}) at the crystal momenta corresponding to the high symmetry points in the $J_1-J_2-\delta$ model BZ shown in Fig. \ref{Fig:J1J2delta_BZ}. The crystal momenta $\vec{q}=(0,0)$, $\vec{q}=(\pi,0)$ and $\vec{q}=(\pi,\pi)$ correspond to $\Gamma$, $X$ and $M$ high symmetry points, respectively, and $\vec{q}=(\pm\pi/2,\pi/2)$ to $\Sigma$ and $\Sigma'$ points. $P$ denotes the spin parity eigenvalue ($P$ is $+1$ or $-1$) in the $S_z^{tot}=0$ sector.
}
\end{figure}

We further discuss the lowest energy excitations in each symmetry sector. The results for the $L\times L\times 4$ clusters with $2L=4n$ and PBC are shown in Fig. \ref{Fig:En}. The lowest energy excitations for the clusters with $2L=4n+2$ and PBC show the same overall structure since the clusters are characterized by the same order parameters as $2L=4n$ clusters although with the reduced value. As it can be seen from the Fig. \ref{Fig:En} the lowest energy excitations are in the $S_z^{tot}=0$ sector and correspond to the irrep $B_1$ with spin parity $P=1$ and crystal momenta $\vec{q}=(0,0)$ and $\vec{q}=(\pi,\pi)$ ($\Gamma$ and $M$ high symmetry points in the BZ) and to the irrep $B_2$ with spin parity $P=1$ and crystal momentum $\vec{q}=(\pi,0)$ ($X$ high symmetry point in the BZ). These excitations correspond to the Goldstone modes in the found bond-nematic state.

The Goldstone modes related to fluctuations of the bond-nematic order parameter are massless quadrupolar excitations of the bond-nematic state and correspond to the rotations of the nematic director that represents the preferred average orientation. Since we find $S_z^{tot}=0$ for these modes they are related to longitudinal fluctuations with $\Delta S_z^{tot}=0$. Also, the Goldstone modes share the same spin parity $P=1$ as the ground state since small oscillations that correspond to these modes only infinitesimally rotate direction of the bond-nematic axis (director) and do not flip the global direction of the axis. 

Unlike the $A_1$ irrep that is fully symmetric (trivial) under all point group operations, irreps $B_1$ and $B_2$ are not symmetric under $C_4$ ($\pi/2$) rotation. In addition, irrep $B_1$ is not symmetric with respect to the diagonal reflection symmetry $\sigma_d$ and irrep $B_2$ with respect to the vertical reflection symmetry $\sigma_v$. This can be readily seen from the character table in Fig. \ref{Fig:C4v} where characters corresponding to $C_4$ rotation and reflection symmetries are $-1$ and not $+1$. If the Goldstone modes break $C_4$ and reflection symmetries the underlying ground state also breaks the same symmetries since the Goldstone modes are a direct consequence of the ground state spontaneous symmetry breaking. In addition, the ground state is evidently degenerate and the ground state manifold contains the state that corresponds to the non-trivial $A_2$ irrep at the crystal momentum $\vec{q}=(\pi,0)$ ($X$ high symmetry point in the BZ) as demonstrated in Fig. \ref{Fig:En} (d). The irrep $A_2$ breaks diagonal $\sigma_d$ and vertical $\sigma_v$ reflection symmetries. This is consistent with existence of chiral bond-nematic order with coexisting VBS order. It further shows that the underlying VBS order is columnar-like order rather than the plaquette VBS that maintains four-fold rotational symmetry about the center of the plaquettes and also reflection symmetries. We note that the same values for the order parameters are obtained for the degenerate irrep $A_2$ with crystal momentum $\vec{q}=(\pi,0)$ as for the irrep $A_1$ with crystal momentum $\vec{q}=(0,0)$.

The triplon-like energy levels, with $S_z^{tot}=\pm 1$ or $S_z^{tot}=0$ and spin parity $P=-1$, show the same energy splitting pattern as for the altermagnetic regime (Fig. \ref{Fig:En}). Namely splitting appears between the energy levels corresponding to irreps $A_1$ and $B_1$ ($A_2$ and $B_2$) where maximum splitting can be seen by comparing energy levels at $\vec{q}=(\pi/2,\pi/2)$ and $\vec{q}=(-\pi/2,\pi/2)$ ($\Sigma$ and $\Sigma'$ points in the BZ), and between the energy levels corresponding to irreps $A_1$ and $A_2$ ($B_1$ and $B_2$) where maximum splitting can be seen by comparing energy levels at $\vec{q}=(0,0)$ and $\vec{q}=(\pi,0)$. Similarly as for the magnon modes in the altermagnetic regime, this indicates opposite chiralities for the triplon-like modes that exhibit chiral splitting of the energy levels. In addition we find splitting between the energy levels for $S_z^{tot}= 1$, $S_z^{tot}=-1$  at all crystal momenta in the BZ that we have considered which indicates broken $\mathbb{Z}_2$ spin inversion symmetry ($S_z^{tot}\rightarrow -S_z^{tot}$). This is consistent with the non-zero value of the squared bond-nematic order parameter $ (O^{yz})^2$ in the ground state shown in Fig. \ref{Fig:Q_scaling}. Namely, the off-diagonal $\hat{O}^{yz}$ (Eq. (\ref{eq:N_op})) operator is an odd-parity operator under $\mathbb{Z}_2$ spin inversion transformation $S_z^{tot}\rightarrow -S_z^{tot}$ and its non-zero expectation value signals $\mathbb{Z}_2$ spin inversion symmetry breaking and splitting between the energy levels with $S_z^{tot}=+1$ and $-1$ since degeneracy of $S_z^{tot}=\pm 1$ energy levels is no longer protected by the $\mathbb{Z}_2$ spin inversion symmetry. Additionally off-diagonal $\hat{O}^{yz}$ (Eq. (\ref{eq:N_op})) mixes different $S_z$ components and its non-zero value implies spontaneous breaking of the $U(1)$ spin rotation symmetry around the $z$-axis. The two-magnon condensation related to the magnon pairing operator $\hat{S}_i^+\hat{S}_j^{+}$ also breaks $U(1)$ spin rotation symmetry, as discussed earlier in this section.

In summary, we find that low energy spectrum contains bond-nematic Goldstone modes and chiral triplon-like excitations exhibiting broken $U(1)$ spin rotation and $\mathbb{Z}_2$ spin inversion symmetries and chiral splitting similar to the chiral splitting in the altermagnetic regime. The low energy spectrum therefore clearly reflects exceptional complexity of the magnetically disordered phase with coexisting bond-nematic and SPT VBS order. 

\section{Conclusions}
\label{sec:Conclusions}
We have studied spin-$1/2$ square lattice $J_1-J_2-\delta$ model using novel ML approach that combines symmetry enhanced NQS ans\"atze and VMC based on SR method. Our results confirm that in the regime of low frustration the model supports an altermagnetic ground state with characteristic chiral splitting of the magnon modes.  In the highly frustrated regime with relatively large exchange interaction modulation parameter $\delta$ we find an intriguing magnetically disordered phase that hosts coexisting bond-nematic and SPT VBS orders that appear as a consequence of melting of the altermagnetic order when quantum fluctuations are increased due to increased frustration. Our study therefore presents an important step in identifying exotic phases of matter that can appear in the vicinity of the altermagnetic order. As demonstrated in our study in addition to possible spin liquid phases a bond-nematic phase can also appear. The bond-nematic phase with coexisting SPT VBS order that we have identified clearly reflects exceptional complexity that appears due to interplay of symmetry and topology. Its low energy spectrum features bond-nematic Goldstone modes and chiral triplon-like modes that exhibit chiral splitting similar to the chiral splitting for the magnon modes in the altermagnetic regime.

The complexity of phases found so far in the phase diagram of the $J_1-J_2-\delta$ model suggests possible unconventional quantum phase transitions and DQC points. Studying the nature of the phase transitions is a direction of our future research. Further interesting future research directions are examining nature of the superconductivity that appears in the model upon doping within fermionic NQS approach and studying magnetically disordered phases for spin-$1$ counterpart of the spin-$1/2$ $J_1-J_2-\delta$ model. 

\begin{acknowledgments} 
It is a pleasure to thank Anders Sandvik for fruitful discussions and facilitating access to Boston University's Resaerch Computing Services. This work is supported by the Singapore Ministry of Education (MOE) Academic Research Fund 
MOE-MOET32023-0003 and MOE-T2EP50223-0019. We would also like to acknowledge the NTU High Performance Computing Centre and the Shared Computing Cluster managed by Boston University’s Research Computing Services for providing computing resources, facilities, and services that have contributed to this work.
\end{acknowledgments}


\begin{thebibliography}{99}

\bibitem{Smejkal}L. \v{S}mejkal, J. Sinova, and T. Jungwirth, \emph{Beyond Conventional Ferromagnetism and Antiferromagnetism: A Phase with Nonrelativistic Spin and Crystal Rotation Symmetry}, Phys. Rev. X \textbf{12}, 031042 (2022). 
\bibitem{Smejkal2}L. \v{S}mejkal, J. Sinova, and T. Jungwirth, \emph{Emerging Research Landscape of Altermagnetism}, Phys. Rev. X \textbf{12}, 040501 (2022).
\bibitem{Mazin}I. Mazin, \emph{Altermagnetism then and now}, Physics \textbf{17}, 4 (2024).
\bibitem{Mazin2}I. Mazin and The PRX Editors, \emph{Editorial: Altermagnetism—A new punch line of fundamental magnetism}, Phys. Rev. X \textbf{12}, 040002 (2022).
\bibitem{Jungwirth}T. Jungwirth, R. M. Fernandes, E. Fradkin, A. H. MacDonald, J. Sinova, and L. \v{S}mejkal, \emph{Altermagnetism: An unconventional spin-ordered phase of matter}, Newton \textbf{1}, 100162 (2025).
\bibitem{Jungwirth2}T. Jungwirth, J. Sinova, P. Wadley, D. Kriegner, H. Reichlov\'a, F. Krizek, H. Ohno, and L. \v{S}mejkal, \emph{Altermagnetic spintronics}, arXiv:2508.09748.
\bibitem{Song}C. Song, H. Bai, Z. Zhou \emph{et al.}, \emph{Altermagnets as a new class of functional materials}, Nat. Rev. Mater. \textbf{10}, 473–485 (2025). 
\bibitem{Smejkal3}L. \v{S}mejkal, A. Marmodoro, K.-H. Ahn, R. Gonz\'alez-Hern\'andez, I. Turek, S. Mankovsky, H. Ebert, S. W. D'Souza, O. \v{S}ipr \emph{et al.}, \emph{Chiral Magnons in Altermagnetic $RuO_2$}, Phys. Rev. Lett. \textbf{131}, 256703 (2023).
\bibitem{Chumak}A. Chumak, V. Vasyuchka, A. Serga \emph{et al.}, \emph{Magnon spintronics}, Nature Phys. \textbf{11}, 453–461 (2015).
\bibitem{Bychkov}Y. A. Bychkov, and E. I. Rashba, \emph{Properties of a 2D electron gas with lifted spectral degeneracy}, JETP Lett. \textbf{39}, 78–81 (1984).
\bibitem{Dresselhaus}G. Dresselhaus, \emph{Spin-Orbit Coupling Effects in Zinc Blende Structures}, Phys. Rev. \textbf{100}, 580 (1955).
\bibitem{Beida}W. Beida, E. \c{S}a\c{s}\i o\v{g}lu, C. Friedrich \emph{et al.}, \emph{Chiral split magnons in metallic g-wave altermagnets: insights from many-body perturbation theory}, npj Quantum Mater. \emph{10}, 97 (2025). 
\bibitem{Liu}Y. Liu, S. Shao, S. He, Z. Y. Xie, J.-W. Mei, H.-G. Luo, and J. Zhao, \emph{Quantum dynamics in a spin-$1/2$ square lattice $J_1-J_2-\delta$ altermagnet}, Phys. Rev. B \textbf{111}, 245117 (2025).
\bibitem{Chen2}H. Chen, G. Duan, C. Liu, Y. Cui, W. Yu, Z. Y. Xie, and R. Yu, \emph{Spin excitations of the Shastry-Sutherland model - altermagnetism and deconfined quantum criticality}, arXiv:2411.00301.
\bibitem{Xie}Y. Xie, D. Wang, C. Li, X. Shen, and J. Zhang, \emph{A General Theory of Chiral Splitting of Magnons in Two-Dimensional Magnets}, arXiv:2601.15031.
\bibitem{Attias}L. Attias, A. Levchenko, and M. Khodas, \emph{Intrinsic anomalous Hall effect in altermagnets}, Phys. Rev. B \textbf{110}, 094425 (2024).
\bibitem{Sheoran}S. Sheoran, and P. Dev, \emph{Spontaneous anomalous Hall effect in two-dimensional altermagnets}, Phys. Rev. B \textbf{111}, 184407 (2025).
\bibitem{Yi}X.-J. Yi, Y. Mao, X. Lu, and Q.-F. Sun, \emph{Spin splitting Nernst effect in altermagnets}, Phys. Rev. B \textbf{111}, 035423 (2025).
\bibitem{Weissenhofer} M. Wei{\ss}enhofer, M. S. Mrudul, S. Mankovsky, and P. M. Oppeneer, \emph{Magnon orbital Nernst effect in altermagnets}, npj Quantum Mater. \textbf{11}, 25 (2026). 
\bibitem{Ogawa}Y. Ogawa, and S. Hayami, \emph{Nonlinear Piezomagnetic Effects in g-wave Altermagnets}, J. Phys. Soc. Jpn. \textbf{94}, 063704 (2025). 
\bibitem{Bell}B. Bell, and J. W. F. Venderbos, \emph{Orbital piezomagnetic polarizability of pure insulating altermagnets in two dimensions}, arXiv:2602.10076. 
\bibitem{Mazin3} I. I. Mazin, \emph{Notes on altermagnetism and superconductivity},  AAPPS Bull. \textbf{35}, 18 (2025). 
\bibitem{Liu2}Z. Liu, H. Hu, and X.-J. Liu, \emph{Altermagnetism and Superconductivity: A Short Historical Review}, arXiv:2510.09170. 
\bibitem{Sobral}J. A. Sobral, S. Mandal, and M. S. Scheurer, \emph{Fractionalized altermagnets: From neighboring and altermagnetic spin liquids to spin-symmetric band splitting}, Phys. Rev. Research \textbf{7}, 023152 (2025). 
\bibitem{Zhu}J.-X. Zhu, R. Yu, H. Wang, L. L. Zhao, M. D. Jones, J. Dai, E. Abrahams, E. Morosan, M. Fang, and Q. Si, \emph{Band Narrowing and Mott Localization in Iron Oxychalcogenides $La_2O_2Fe_2O(Se,S)_2$}, Phys. Rev. Lett. \textbf{104}, 216405 (2010).
\bibitem{Das} P. Das, V. Leeb, J. Knolle, and M. Knap, \emph{Realizing Altermagnetism in Fermi-Hubbard Models with Ultracold Atoms}, Phys. Rev. Lett. \textbf{132}, 263402 (2024).
\bibitem{Ma} H.-Y. Ma, M. Hu, N. Li, J. Liu, W. Yao, J.-F. Jia, and J. Liu, \emph{Multifunctional antiferromagnetic materials with giant piezomagnetism and noncollinear spin current}, Nat. Commun. \textbf{12}, 2846 (2021).
\bibitem{Roth}C. Roth, A. Szab\'{o}, and A. MacDonald, \emph{High-accuracy variational Monte Carlo for frustrated magnets with deep neural networks}, Phys. Rev. B \textbf{108}, 054410 (2023).
\bibitem{Morita}S. Morita, R. Kaneko, and M. Imada, \emph{Quantum Spin Liquid in Spin 1/2 $J_1–J_2$ Heisenberg Model on Square Lattice: Many-Variable Variational Monte Carlo Study Combined with Quantum-Number Projections}, J. Phys. Soc. Jpn. \textbf{84}, 024720 (2015).  
\bibitem{Wang}L. Wang, and A. W. Sandvik, \emph{Critical Level Crossings and Gapless Spin Liquid in the Square-Lattice Spin-1/2 $J_1-J_2$ Heisenberg Antiferromagnet}, Phys. Rev. Lett. \textbf{121}, 107202 (2018).
\bibitem{Ferrari}F. Ferrari, and F. Becca, \emph{Gapless spin liquid and valence-bond solid in the $J_1-J_2$ Heisenberg model on the square lattice: Insights from singlet and triplet excitations}, Phys. Rev. B \textbf{102}, 014417 (2020). 
\bibitem{Nomura}Y. Nomura, and M. Imada, \emph{Dirac-Type Nodal Spin Liquid Revealed by Refined Quantum Many-Body Solver Using Neural-Network Wave Function, Correlation Ratio, and Level Spectroscopy}, Phys. Rev. X \textbf{11}, 031034 (2021).
\bibitem{Liu3}W.-Y. Liu, S.-S. Gong, Y.-B. Li, D. Poilblanc, W.-Q. Chen, and Z.-C. Gu, \emph{Gapless quantum spin liquid and global phase diagram of the spin-1/2 square antiferromagnetic Heisenberg model}, 	Science Bulletin \emph{67}, 1034-1041(2022).
\bibitem{Feuerpfeil}A. Feuerpfeil, L. Shackleton, A. Maity, R. Thomale, S. Sachdev, and  Y. Iqbal, \emph{Unifying Dirac Spin Liquids on Square and Shastry-Sutherland Lattices via Fermionic Deconfined Criticality}, arXiv:2601.19980. 
\bibitem{Maity}A. Maity, A. Feuerpfeil, R. Thomale, S. Sachdev, and Y. Iqbal, arXiv:2603.15745. 
\bibitem{Zou}H. Zou, F. Yang, and W. Ku, \emph{Nearly degenerate ground states of a checkerboard antiferromagnet and their bosonic interpretation}, Sci. China Phys. Mech. Astron. \textbf{67}, 217211 (2024). 
\bibitem{Duric}T. \DJ uri\'c, J. H. Chung, B. Yang, and P. Sengupta, \emph{Spin-1/2 Kagome Heisenberg Antiferromagnet: Machine Learning Discovery of the Spinon Pair-Density-Wave Ground State}, Phys. Rev. X \textbf{15}, 011047 (2025).
\bibitem{Duric2}T. \DJ uri\'c, and P. Sengupta, arXiv:2512.11670.
\bibitem{Carleo}G. Carleo, K. Choo, D. Hofmann, J. E. T. Smith, T. Westerhout, F. Alet, E. J. Davis, S. Efthymiou, I. Glasser, S.-H. Lin, M. Mauri, G. Mazzola, C. B. Mendl, E.
van Nieuwenburg, O. O'Reilly, H. Th\'eveniaut, G. Torlai, and A. Wietek, NetKet: \emph{A Machine Learning Toolkit for Many-Body Quantum Systems} , SoftwareX \textbf{10} , 100311
(2019).
\bibitem{Vicentini}F. Vicentini, D. Hofmann, A. Szab\'o, D. Wu, C. Roth, C.Giuliani, G. Pescia, J. Nys, V. Vargas-Calder\'on, N. Astrakhantsev, and G. Carleo, NetKet 3: Machine Learning
Toolbox for Many-Body Quantum Systems , SciPost Phys. Codebases \textbf{7} (2022); https://www.netket.org/.
\bibitem{Frostig}R. Frostig, M. J. Johnson, and C. Leary, \emph{Compiling machine learning programs via high-level tracing}, Systems for Machine Learning (2018).
\bibitem{Heek} J. Heek, A. Levskaya, A. Oliver, M. Ritter, B. Rondepierre, A. Steiner, and M. van Zee, \emph{Flax: A neural network library and ecosystem for JAX},
http://github.com/google/flax (2020).
\bibitem{Hessel}M. Hessel, D. Budden, F. Viola, M. Rosca, E. Sezener, and T. Hennigan, \emph{Optax: Composable gradient transformation and optimisation in JAX}, http://github.com/deepmind/optax (2020).
\bibitem{Sorella1}S. Sorella, \emph{Green Function Monte Carlo with Stochastic Reconfiguration}, Phys. Rev. Lett. \textbf{80} , 4558 (1998).
\bibitem{Sorella2}S. Sorella, \emph{Generalized Lanczos algorithm for variational quantum Monte Carlo}, Phys. Rev. B \textbf{64} , 024512 (2001).
\bibitem{Rende}R. Rende, L. L. Viteritti, L. Bardone, F. Becca, and S. Goldt, \emph{A simple linear algebra identity to optimize large-scale neural network quantum states}, Commun. Phys. \textbf{7}, 260 (2024).
\bibitem{Chen}A. Chen, and M. Heyl, \emph{Empowering deep neural quantum states through efficient optimization}, Nat. Phys. \textbf{20}, 1476–1481 (2024).
\bibitem{Chojnacki}L. Chojnacki, R. Pohle, H. Yan, Y. Akagi, and N. Shannon, \emph{Gravitational wave analogs in spin nematics and cold atoms}, Phys. Rev. B \textbf{109}, L220407 (2024).
\bibitem{Viteritti}L. L. Viteritti, R. Rende, S. Sachdev, and G. Carleo, \emph{Approaching the Thermodynamic Limit with Neural-Network Quantum States}, arXiv:2602.02665. 
\bibitem{Moss}M. S. Moss, R. Wiersema, M. Hibat-Allah, J. Carrasquilla, and R. G. Melko, \emph{Leveraging recurrence in neural network wavefunctions for large-scale simulations of Heisenberg antiferromagnets on the square lattice}, Phys. Rev. B \textbf{112}, 134450 (2025).
\bibitem{Anderson}P. W. Anderson, \emph{An Approximate Quantum Theory of the Antiferromagnetic Ground State}, Phys. Rev. \textbf{86}, 694 (1952).
\bibitem{Tasaki}H. Tasaki, \emph{Long-Range Order, "Tower" of States, and Symmetry Breaking in Lattice Quantum Systems}, J. Stat. Phys. \textbf{174}, 735–761 (2019).
\bibitem{Wietek}A. Wietek, M. Schuler, and A. M. L\"auchli, \emph{Studying Continuous Symmetry Breaking using Energy Level Spectroscopy}, arXiv:1704.08622. 
\bibitem{Toth}S. Toth and B. Lake, \emph{Linear spin wave theory for single-Q incommensurate magnetic structures}, J. Phys.: Condens. Matter \textbf{27}, 166002 (2015).
\bibitem{Senthil}T. Senthil, A. Vishwanath, L. Balents, S. Sachdev, and M. P. A. Fisher, \emph{Deconfined Quantum Critical Points}, Science \textbf{303}, 5663, 1490-1494 (2004).
\bibitem{Senthil2}T. Senthil, L. Balents, S. Sachdev, A. Vishwanath, and M. P. A. Fisher, \emph{Deconfined Criticality Critically Defined}, J. Phys. Soc. Jpn. \textbf{74}, pp. 1-9 (2005)
\bibitem{Cui}Y. Cui, R. Yu and W. Yu, \emph{Deconfined Quantum Critical Point: A Review of Progress}, Chinese Phys. Lett. \textbf{42}, 047503 (2025).
\bibitem{Zhitomirsky} M. E. Zhitomirsky and H. Tsunetsugu, \emph{Magnon pairing in quantum spin nematic}, EPL \textbf{92}, 37001 (2010).
\bibitem{Wang3}Z. Wang, and C. D. Batista, \emph{Dynamics and Instabilities of the Shastry-Sutherland Model}, Phys. Rev. Lett. \textbf{120}, 247201 (2018).
\bibitem{Yokoyama}Y. Yokoyama and C. Hotta, \emph{Spin nematics next to spin singlets}, Phys. Rev. B \textbf{97}, 180404(R) (2018).
\bibitem{Momoi}T. Momoi, P. Sindzingre, and K. Kubo, \emph{Spin Nematic Order in Multiple-Spin Exchange Models on the Triangular Lattice}, Phys. Rev. Lett. \textbf{108}, 057206 (2012). 
\bibitem{Haghshenas}R. Haghshenas, A. Langari and A. T. Rezakhani, \emph{Symmetry fractionalization: symmetry-protected topological phases of the bond-alternating spin-1/2 Heisenberg chain}, J. Phys.: Condens. Matter \textbf{26}, 456001 (2014). 
\bibitem{Wang2}Z. Wang, L. Lu, S.-Q. Ning, Z. Liu, Y.-C. Wang, Z. Yan, and W. Guo, \emph{Detecting underlying symmetry-protected topological phases via strange correlators and edge engineering}, Phys. Rev. B \textbf{113}, 054405 (2026).
\bibitem{Shannon}N. Shannon, T. Momoi, and P. Sindzingre, \emph{Nematic Order in Square Lattice Frustrated Ferromagnets}, Phys. Rev. Lett. \textbf{96}, 027213 (2006).

\end{thebibliography}
\end{document}